\documentclass[10pt]{article}

\usepackage{amsmath}
\usepackage{array}
\usepackage{appendix}
\usepackage{graphicx}
\usepackage{amsfonts}
\usepackage{amssymb}
\usepackage{mathrsfs}
\usepackage{yfonts}
\usepackage{euscript}
\usepackage{upgreek}
\usepackage{slantsc}
\usepackage{calligra}
\usepackage[T1]{fontenc}
\usepackage{epsf}
\usepackage{latexsym}

\usepackage{tipa}

\def\be{\begin{equation}}
\def\ee{\end{equation}}
\def\beq{\begin{equation}}
\def\eeq{\end{equation}}
\def\bea{\begin{eqnarray}}
\def\eea{\end{eqnarray}}

\def\ni{\noindent}
\def\foo{\footnote}

\def\hat{\widehat}

\def\!{\hspace{-1.6667em}}

\def\mN{\mbox{N}}

\def\mV{\mbox{V}}

\def\md{\mbox{d}} 
\def\me{\mbox{e}}

\def\ml{\mbox{l}}   
\def\mm{\mbox{m}}

\def\urho{{\underline{\rho}}}
\def\upi{{\underline{\pi}}}
\def\usrho{{\underline{\mbox{\scriptsize $\rho$}}}}

\def\brho{\mbox{\boldmath$\rho$}}          
\def\bpi{\mbox{\boldmath$\pi$}}            %

\def\bupSigma{\mbox{\boldmath$\Sigma$}}                 

\def\fA{\mbox{\sffamily A}}
\def\fB{\mbox{\sffamily B}}
\def\fC{\mbox{\sffamily C}}

\def\fQ{\mbox{\sffamily Q}}

\def\r{\underline{r}}

\def\q{\underline{q}}

\def\bl{\mbox{\bf l}}

\def\bp{\mbox{\bf p}}

\def\bG{\mbox{{\bf G}}}

\def\bm{\mbox{{\bf m}}}

\def\bq{\mbox{{\bf q}}}

\def\bz{\mbox{{\bf z}}}

\def\scK{\mbox{\scriptsize ${\cal K}$}}

\def\bigr{\mbox{\boldmath$\mathfrak{R}$}}

\def\Fr{\mbox{\Large $\mathfrak{r}$}}

\def\FP{\mbox{\Large $\mathfrak{p}$}}
\def\FrQ{\mbox{\Large $\mathfrak{q}$}}
\def\FrN{\mathfrak{N}}                            
 
\def\FrA{\mbox{\boldmath$\mathfrak{A}$}} 

\def\FrH{\mbox{\boldmath$\mathfrak{H}$}}          
\def\FrV{\mbox{\Large\boldmath$\mathfrak{v}$}}    

\def\FrC{\mbox{\Large $\mathfrak{c}$}}

\def\sFG{\mbox{$\mathfrak{g}$}}

\def\FrS{\mbox{\Large $\mathfrak{s}$}}

\def\FrG{\mbox{\Large $\mathfrak{g}$}}  
\def\nFrG{\mbox{\large $\mathfrak{g}$}}
\def\nFrV{\mbox{\large $\mathfrak{v}$}}  

\def\nFrS{\mbox{\large $\mathfrak{s}$}}  
\def\Frg{\mbox{\normalsize $\mathfrak{g}$}}                                  
\def\Frk{\mbox{\scriptsize $\mathfrak{K}$}}                       
\def\Frh{\mbox{$\mathfrak{h}$}} 

\def\FrO{\mbox{\Large $\mathfrak{o}$}}
   
\def\FA{\mbox{\Large $\mathfrak{a}$}}

\def\FS{\mbox{\LARGE\tt s}}                         

\def\sg{\mbox{\scriptsize g}} 
\def\sh{\mbox{\scriptsize h}} 
\def\si{\mbox{\scriptsize i}}

\def\sn{\mbox{\scriptsize n}}

\def\sr{\mbox{\scriptsize r}}
\def\sss{\mbox{\scriptsize s}}  

\def\su{\mbox{\scriptsize u}}
\def\sv{\mbox{\scriptsize v}}

\def\sy{\mbox{\scriptsize y}}

\def\sC{\mbox{\scriptsize C}}

\def\sT{\mbox{\scriptsize T}}

\def\sfh{\mbox{\sffamily{\scriptsize h}}}     
\def\sfk{\mbox{\sffamily{\scriptsize k}}}     
\def\sfo{\mbox{\sffamily{\scriptsize o}}}     
\def\sfp{\mbox{\sffamily{\scriptsize p}}}     
\def\sfq{\mbox{\sffamily{\scriptsize q}}}     
\def\sfr{\mbox{\sffamily{\scriptsize r}}}     
\def\sfs{\mbox{\sffamily{\scriptsize s}}}     

\def\sfA{\mbox{\sffamily{\scriptsize A}}}      
\def\sfB{\mbox{\sffamily{\scriptsize B}}}      
\def\sfC{\mbox{\sffamily{\scriptsize C}}}      
\def\sfE{\mbox{\sffamily{\scriptsize E}}}      
\def\sfG{\mbox{\sffamily{\scriptsize G}}}      
\def\tfA{\mbox{\sffamily{\tiny A}}}

\def\B{\underline{B}}

\def\K{Kucha\v{r} }

\def\pa{\partial}
\def\d{\textrm{d}}

\def\5Star{\mbox{\Large$\star$}}              
\def\cr{\mbox{\scriptsize{\bf $\mbox{ } \times \mbox{ }$}}}

\def\shortcr{\mbox{\scriptsize{\bf $\, \times \,$}}}

\def\sumi2{\sum\mbox{}_{\mbox{}_{\mbox{\scriptsize $i$=1}}}^2}
\def\sumi3{\sum\mbox{}_{\mbox{}_{\mbox{\scriptsize $i$=1}}}^3}

\def\sumAn{\sum\mbox{}_{\mbox{}_{\mbox{\scriptsize $A$=1}}}^{n}}
\def\sumABn{\sum\mbox{}_{\mbox{}_{\mbox{\scriptsize $A,B$=1}}}^{n}}
\def\sumCDn{\sum\mbox{}_{\mbox{}_{\mbox{\scriptsize $C,D$=1}}}^{n}}

\def\sumABcycles3{\sum\mbox{}_{\mbox{}_{\mbox{\scriptsize cycles  $A,B$=1}}}^{3}}
\def\sumCDcycles3{\sum\mbox{}_{\mbox{}_{\mbox{\scriptsize cycles  $C,D$=1}}}^{3}}

\def\sumIN{\sum\mbox{}_{\mbox{}_{\mbox{\scriptsize $I$=1}}}^{N}}
\def\sumJN{\sum\mbox{}_{\mbox{}_{\mbox{\scriptsize $J$=1}}}^{N}}

\def\sumj3{\sum\mbox{}_{\mbox{}_{\mbox{\scriptsize $j$=1}}}^3}
\def\sumk3{\sum\mbox{}_{\mbox{}_{\mbox{\scriptsize $k$=1}}}^3}

\def\timesIN{\bigtimes\mbox{}_{\mbox{}_{\mbox{\scriptsize $I$=1}}}^{N}}
\def\bigtimes{\mbox{\Large $\times$}}
\def\sumG{\sum\mbox{}_{\mbox{}_{\mbox{\scriptsize g $\in \sFG$}}}}                           

\begin{document}

\begin{titlepage}

\begin{center}


{\bf \LARGE Six New Mechanics corresponding to further Shape Theories}

\vspace{.15in}

{\large \bf Edward Anderson} 

\vspace{.15in}

\large {\em DAMTP, Centre for Mathematical Sciences, Wilberforce Road, Cambridge CB3 OWA.  } \normalsize

\end{center}

\begin{abstract}

A suite of relational notions of shape are presented at the level of configuration space geometry, with corresponding new theories of shape mechanics and shape statistics.
These further generalize two quite well known examples: --1) Kendall's (metric) shape space with his shape statistics and Barbour's mechanics thereupon.
0) Leibnizian relational space alias metric scale-and-shape space to which corresponds Barbour--Bertotti mechanics. 
This paper's new theories include, using the invariant and group namings, 
1) {\it Angle}              alias {\it conformal shape mechanics}.  
2) {\it Area ratio}         alias {\it affine shape mechanics}.
3) {\it Area}               alias {\it affine scale-and-shape mechanics}.
1) to 3) rest respectively on angle space, area-ratio space, and area space configuration spaces.  
%
%
%
Probability and statistics applications are also pointed to in outline.

4) Various supersymmetric counterparts of --1) to 3) are considered. 
Since supergravity differs considerably from GR-based conceptions of Background Independence, some of the new supersymmetric shape mechanics are compared with both.
These reveal compatibility between supersymmetry and GR-based conceptions of Background Independence, at least within these simpler model arenas.  
 
\end{abstract}

\end{titlepage}

\section{Introduction}\label{Introduction}

Newtonian Mechanics, and the Newtonian paradigm of Physics more generally, are based on absolute space and time. 
The immovable external character of absolute space and time led to these being opposed by {\it relationalists}, most notably by Leibniz \cite{L} and Mach \cite{M}.  
On the other hand, the Newtonian paradigm of Physics sufficed to explain humankind's observations 
of nature until the end of the 19th century.

Indeed, a satisfactory relational alternative to the foundations of Mechanics was not found until 1982 by Barbour and Bertotti \cite{BB82}. 
This is a theory in which Euclidean transformations are held to be physically irrelevant.
The next RPM was not formulated until 2003; it is Barbour's `mechanics of pure shape' \cite{B03}; in this case, it is similarity transformations which are held to be physically irrelevant.
It is useful to term this a Shape Mechanics, and the preceding a Scale-and-Shape Mechanics.  
I then considered what the configuration spaces are for these two theories.
It turns out that this Shape Mechanics' configuration spaces (`shape spaces') are simpler, 
with the corresponding Shape-and-Scale Mechanics' configuration spaces then being the cones over these \cite{Cones}.  
E.g. the shape spaces for $N$ particles in 1-$d$ are $\{N - 2\}$-spheres $\mathbb{S}^{N - 2}$, and those in 2-$d$ are complex projective spaces $\mathbb{CP}^{N - 2}$; 
furthermore for 3 particles in 2-$d$, $\mathbb{CP}^1 = \mathbb{S}^2$: the {\it shape sphere} of all possible triangular shapes.  
Furthermore, these shape spaces turned out to have already arisen in Kendall's studies of the Geometry, Probability and Statistics of shapes (\cite{Kendall84, Kendall, Small}).
This is a very useful interdisciplinary connection which I pointed out in the Theoretical Physics literature in \cite{FORD}.  
I further studied these two RPMs in \cite{FileR}, eventually summarizing the key properties of their configuration spaces within my review \cite{AConfig} on 
configuration spaces in Theoretical Physics more generally.  

\mbox{ }

\ni RPMs have a number of foundationally valuable applications in Theoretical Physics \cite{FileR}, including the following.

\mbox{ } 

\ni 1) RPMs have a number of features in common with GR as viewed as a dynamical system \cite{ADM, Battelle, San-1, FileR}. 
In particular, they have an energy constraint analogous to the GR Hamiltonian constraint; both are quadratic in the momenta.
They also have constraints linear in the momenta (and first-class) which are analogues of the GR momentum constraint.

\ni 2) RPMs are useful in analyzing which aspects of Background Independence GR possesses.
Difficulties with these then become facets of the notorious Problem of Time in Quantum Gravity \cite{Kuchar92, EOT, RWR, KieferBook, FileR, ABeables, BI, APoT3, ABook}.  
Each constraint that is quadratic in the momenta can be taken to arise from the corresponding theory's reparametrization invariance, 
which implements primary-level timelessness for whole-universe models.
Each constraint that is linear in the momenta (and first-class) can be taken to arise from a group $\FrG$ of physically irrelevant transformations acting on the configuration space $\FrQ$ of the theory.  
In the case of GR, $\FrG = Diff(\bupSigma)$: the spatial diffeomorphisms on the 3-space of fixed topology $\bupSigma$, 
whereas $\FrQ = \mbox{Riem}(\bupSigma)$: the spatial 3-metrics on $\bupSigma$.  
The corresponding quotient space of these is Wheeler's $\mbox{Superspace}(\bupSigma) = \mbox{Riem}(\bupSigma)/Diff(\bupSigma)$ \cite{DeWitt67, GR-Config}; 
RPMs then offer model arenas of other such quotient spaces as well: shape spaces or relational spaces alias scale-and-shape spaces.  
Some of these are then more closely analogous to further GR configuration spaces such as conformal superspace \cite{GR-Config2}; 
see Appendix C for more detailed comparison between RPM and GR configuration spaces.  

\ni 3) I also applied \cite{AKendall} Kendall's own case of Shape Statistics to Timeless Records Theory \cite{EOT}. 
This is one of the various approaches to the Problem of Time; in this case, one sees how far one can get by addressing timeless propositions.
In particular, my application makes it concrete that Shape Statistics provides the machinery requisite for rendering the classical version of Timeless Records Theory mathematically sharp.

\ni 4) RPMs are useful for modelling closed universe and quantum cosmological effects \cite{DeWitt67, FileR}.

\ni 5) RPMs are also useful models \cite{Rovelli, FileR} in the study of geometrical quantization \cite{I84}, and affine quantum geometrodynamics \cite{IK84, Affine}.

\mbox{ } 

\ni The current paper represents a major extension of the scope of RPMs: from one theory three decades ago \cite{BB82}
and a second theory one decade ago \cite{B03} to now presenting a large number of RPM theories, and moreover derived in a systematic manner.
This is based on knowing the standard set of geometries on flat space, standard group theory of the corresponding transformation groups and the interplay between the two \cite{Stillwell}. 
The expansion in scope then rests upon the similarity group $Sim(d)$ being a subgroup of both the affine group $Aff(d)$ and the conformal group $Conf(d)$.
These are in fact two mutually incompatible extensions, so in any given model one has to make a choice between these `apex groups'. 

Each of these `apex groups' has numerous subgroups, including some shared between them. 
While this is elsewhere well-known mathematics, it was not brought into consideration by Barbour, nor by any of the other authors working in this field.
However, upon making various reconceptualizations as laid out in Sec 2, 
one can readily tap into this material to produce more general notions of shape, corresponding notions of shape space and corresponding RPM theories.

Sec 2 covers this expansion at the level of groups, invariants and the correseponding configurations.
In Sec 3, I follow this up by consider the corresponding configuration spaces' geometry. 
In particular, in Sec 2 I provide the {\it minimal relationally nontrivial unit}.
This is concurrently the smallest relationally nontrivial i) whole-universe model, ii) dynamical subsystem and iii) Shape Statistics sampling unit.  
The archetype of minimal relationally nontrivial unit is the relational triangle; e.g. Barbour's seminars have often involved demonstrations involving shuffling wooden triangles. 
Furthermore, the corresponding crucial technical tools are based on knowledge of the topology and geometry of the corresponding configuration space of relational triangles. 
This shape space is the {\it shape sphere}, or a portion thereof, depending on the exact modelling assumptions \cite{AConfig}. 
E.g. Kendall's {\it spherical blackboard} (Fig \ref{Sphe-Chop}.a), which is well known in the Shape Geometry and Shape Statistics literature \cite{Kendall84, Kendall}. 
This consists of  1/3 of a hemisphere corresponding to indistinguishable particles with mirror image triangles identified.
On the other hand, a whole hemisphere is required if the former assumption is dropped, or a whole sphere if both are dropped; 
this furthermore becomes Montgomery's `{\it pair of pants}' \cite{Pants} (in the Celestial Mechanics literature) if both assumptions are dropped and double collisions are excised.
In Sec 3, I then introduce, name and consider the configuration spaces of minimal relationally nontrivial units for a large range of further Shape and Scale-and-Shape Theories.
These are also very much expected to be a technically important nucleus for the corresponding theories of RPM and of (Scale-and-)Shape Statistics, 
for which I also provide matching names.
Due to this, the summary tables in Figs \ref{Tab-1} and \ref{Tab-2} are of substantial importance in all of Shape Theory, Shape Statistics and RPMs alongside its applications 
to understanding the foundations of GR-like theories, of Background Independence and of modelling whole(universe quantum cosmological features.  
This justifies presenting a number of frontiers for subsequent research directions.

Sec 4 outlines known examples of topological and geometrical structure for configuration spaces. 
This illustrates some of the detail that one can eventually expect in the study of the new shape spaces and scale-and-shape spaces introduced in the current paper.
Sec 5 considers the issue of configuration comparers for (Scale-and-)Shape Theories. 
This includes three way comparison between Barbour's approach, Kendall's and DeWitt's -- the last of these having been foundational in the study of GR itself as a dynamical system.

This paper's main worked-out application involves new theories of RPM corresponding to further notions of shape, or of scale-and-shape.  
The new such theories I present in Sec 6 are as follows. 

\mbox{ } 

\ni 1) {\it Conformal Shape Mechanics}: a theory of angles alone which most readily generalizes \cite{B03} 
from a structural perspective due to the continued availability of the Euclidean norm.

\ni 2) {\it Area Mechanics} in 2-$d$ -- tied to {\it equiareal} geometry \cite{Coxeter}. 
I show that Area Mechanics requires its kinetic term be built without evoking the Euclidean norm, which ceases to be a licit structure in this context. 

\ni 3) {\it Area Ratio} alias {\it Affine Shape Mechanics} (then Area Mechanics' alias as {\it Affine Scale-and-Shape Mechanics} becomes clear). 

\mbox{ } 

\ni For each of 2) and 3), I also provide a $d$-dimensional generalization of the underlying shape theory. 
I also then show in Sec 7 that Barbour's Best Matching comparer used in building RPMs continues to thrive in the complex plane $\mathbb{C}$.
I first use this to reformulate \cite{B03} (now renamed as {\sl metric} Shape Mechanics, given all the other theories of Shape Mechanics provided in the current paper!)
I then indicate how the indirect formulation of Mechanics runs into difficulty for the Shape Theory in which the M\"{o}bius group is taken to be physically irrelevant.

Sec 8 then outlines new frontiers of research in Shape Mechanics, with Sec 9 following this up by outlining corresponding new frontiers of research in Shape Statistics. 
This is due to the current paper's range of notions of shape exceeding that considered by Kendall, while remaining amenable to parallel shape space geometry based investigations. 
I.e. I generalize Kendall's own illustrative `standing stones' problem for metric Shape Statistics to affine, conformal and M\"{o}bius Shape Statistics 
and other projective Shape Statistics besides.

Kendall and collaborators' topological, geometrical, probabilistic and statistical work \cite{Kendall84, Kendall89, Small, Kendall}, 
indicates the considerable value of the interdisciplinarity connection \cite{FORD, FileR, AKendall} between this and Metric Shape Mechanics. 
It is furthermore to be expected that such interdisciplinary connections will extend to the range of other shape and scale-and-shape theories considered in this paper. 
Moreover, this Shape Geometry, Probability and Statistics took Kendall's group 20 years to develop for {\sl one} notion of shape; 
thus it should in no way be expected for the current publication to work all of this out. 
The current paper already multiplies by a sizeable factor the number of known Relational Mechanics theories. 
The list of Shape Statistics frontiers and other interdisciplinary applications provided is then to be regarded as a source of future research papers, 
Such interdisciplinarities involving metric (scale-and-)shapes have indeed already been established.  
The current paper then points out that these have conceptual-level analogues for other theories of (scale-and-)shape as outlined in the Frontier sections. 
As well as the above-mentioned interdisciplinarities, further such which are established for shape theories include with 

\mbox{ }

\ni i) Robotics \cite{AG} ((for the reasons given in Secs \ref{Q-Geom} and \ref{MSS}).

\ni ii) Image Analysis \cite{Fitzgibbon} (e.g. shapes photographed from whichever direction). 

\ni iii) Biology \cite{Thompson} (e.g. animal morphology).

\ni iv) Astronomy (e.g. galaxies, CMB patterning, microlensing).

\mbox{ }

\ni Additionally, I do not stop at affine and conformal type theories of Classical Mechanics. 
In Sec 10, I give furthermore the first ever treatment of supersymmetric RPM, alongside frontier questions concerning supershapes.
This is in the context of enlarging $Eucl(d)$ and $Sim(d)$ firstly to $superEucl(d)$ and $superSim(d)$ 
and then onward to two competing super-apex groups: the superconformal and superaffine groups, alongside many other subgroups of these.
Some context and motivation for this development is as follows. 

\mbox{ } 

\ni 1) Whether Relationalism and Background Independence more generally have a similar characterization in Supergravity.
The answer is no (\cite{ABeables, AGates} and Appendix D). 
Due to this, it was a substantial oversight to consider only GR in \cite{Kuchar92, I93} on the assumption that other theories of gravity would be similar in this regard.  

\ni 2) Whether Relationalism is compatible with Supersymmetry \cite{ABeables, AGates}. 
The current paper shows that the answer to this is yes.

\mbox{ } 

\ni I end in Sec 11 by providing the Dirac quantizations for a representative four among the current paper's new RPMs. 
This parallels how e.g. Rovelli \cite{Rovelli} and Smolin \cite{Smolin} quantized \cite{BB82} around a decade after its inception   
(whereas I subsequently considered both Dirac and reduced quantizations of both \cite{BB82} and \cite{B03}).

\mbox{ } 

\ni Appendix A supports the text by outlining the group-based approach to the foundations of flat theories of geometry. 
Appendix B outlines those parts of the theory of Lie algebras and Lie groups that are used in this paper, including supersymmetric counterparts.
Appendix C compares the foundational variety of flat space notions of geometry with that of notions of differential geometry, 
Appendix C also gives an outline of GR's Dirac algebroid of constraints. 
This is provided firstly because RPMs' constraint algebras share a number of features with it.
Secondly, it is provided for for comparison with Appendix D's significantly distinct supergravity algebroid of constraints.
This contrast is of foundational interest as regards canonical quantization, Quantum Gravity and differences in the form in which Background Independence can be manifested.  
Finally, the latter is also contrasted with supersymmetric RPMs' own constraint algebras.

\section{Relational configurations and the corresponding configuration spaces}

First consider configuration space $\FrQ$ \cite{Lanczos}: the space of generalized configurations $Q^{\sfA}$ for a physical system.
The rest of this Sec considers finite flat-space Mechanics examples of configurations and configuration spaces.
N.B. that these in fact promptly come to involve {\it mass-weighted} configurations.

This paper considers the case of point configurations.  
Here $\FrQ = \FrQ(N, d) = \timesIN \FA(d)$, for $\FA(d)$ the absolute space. 
For these, the relevant principles of Configurational Relationalism is that Physics involves not only a $\FrQ$ but also a $\FrG$ of transformations acting upon $\FrQ$ 
that are taken to be physically redundant.
Then two a priori distinct conceptualizations of Configurational Relationalism are then possible for point configurations.  

\mbox{ } 

\ni a) $\FrG$ acts on absolute space $\FA(d)$ (usually $\mathbb{R}^d$).  

\ni b) $\FrG$ acts on configuration space $\FrQ(N, d)$, i.e. acting rather on a material entity of at least some physical content.
%

\mbox{ } 

\ni However, in the case of $\FrQ = \FrQ(N, d) = \timesIN \FA(d)$, the group action takes, particle by particle, the form of a group action on inividual particles in $\FA(d)$.
Due to this, each kind of geometry that can be considered for $\FA(d)$ corresponds to a realtionalism imposed on the whole of $\FrQ(N, d)$.

\ni Thus b) can be resolved by resolving a), which amounts to addressing the groups acting on $\mathbb{R}^d$ has been well-addressed, e.g. in Klein's Erlangen approach to geometry.
In this way, Appendix \ref{Flat-Geom}'s well-known mathematics can be straightforwardly commandeered to settle b) also.  

\mbox{ } 

\ni Some limitations on the choice of $\FrG, \FrQ$ pairs are as follows.  

\mbox{ } 

\ni A) {\it Nontriviality}. $\FrG$ cannot be too large, by a degrees of freedom counting criterion. 
Then using $q := \mbox{dim}(\FrQ)$  and $l :=  \mbox{dim($\FrG$)}$, a theory on $\FrQ/\FrG$ is 
{\it inconsistent}         if $l > q$, 
{\it trivial}              if $l = q$ and 
{\it relationally trivial} if $l = q - 1$.\footnote{This is trivial in the sense that there are no independent degrees of freedom in the principal stratum of the orbit space. 
Moreover, the group action can be such that non-principal strata retain nontrivial dynamical content.
Such non-principal strata occur e.g. in the standard action of $SO(3)$ on $\mathbb{R}^3$ and on the configuration space of $\times_{i = 1}^N \mathbb{R}^3$ of $N$-particle configurations.
This is one way in which the counting argument is `local' rather than`global'.  
Another is how 2 + 1 GR manages to be trivial as regards local degrees of freedom but none the less is capable of possessing global degrees of freedom.}
%
The last of these is because {\it relational nontriviality} requires for one degree of freedom to be expressed in terms of another.
This is as opposed to it being meaningfully expressible in terms of some external or elsewise unphysical `time parameter'.   
 
\ni B) Further {\it structural compatibility} is required.  
\ni A simple example of this is that in considering $d$-dimensional particle configurations, $\FrG$ is to involve the same $d$ (or less, but certainly not more).  

\ni C) A more general structural compatibility criterion is that $\FrG$ is to have a group action.\footnote{A {\it group action} $\alpha$ on a set $X$ is a map 
$\alpha: \nFrG \times X \rightarrow X$ such that 
\ni i) $\{g_1 \circ g_2\} x = g_1 \circ \{g_2 x\}$ (compatibility) and 
\ni ii) $e x = x$                                             $\forall \, x \in X$ (identity).
\ni This terminology continues to apply if $X$ carries further structures.
%
%
Examples include {\it left action}  $g x$, {\it right action} $x g$, and {\it conjugate action} $g x g^{-1}$.  
%
%
By a {\it natural} group action, I just mean one which does not require a choice as to how to relate $X$ and $\nFrG$ due to there being one `obvious way' in which it acts. 
E.g. Perm($X$) acts naturally on $X$, or an $n \times n$ matrix group acts naturally on the corresponding $n$-vectors.
\ni An action is {\it faithful}   if $g_1 \neq g_2$ $\Rightarrow$ $g_1 x \neq g_2 x$ for {\sl some} $x \in X$, 
  whereas it is {\it free} if this is so                                             for {\sl all}  $x \in X$.  
A map is {\it proper} if the inverse map of each compact set is itself compact; a particular subcase of this used below is {\it proper group action}.}
%
A group action's credibility may further be enhanced though its being `natural', 
and some further mathematical advantages are conferred from it being whichever of faithful or free, with the combination of free and proper conferring yet further advantages. 
Sec \ref{Mech} contains a further compatibility condition relevant in the case of Mechanics.

\mbox{ } 

\ni One might additionally wish to choose $\FrG$ for a given $\FrQ$ so as to eliminate {\sl all} trace of any extraneous background entities.\footnote{This 
does not mean that relationalists necessarily discard structures, but rather that they are prepared to consider the outcome of entertaining more minimalist ontologies.}
%
The automorphism group\footnote{A {\it homomorphism} is a map $\mu: \nFrS_1 \rightarrow \nFrS_2$ that is structure-preserving. 
In particular, if a such is invertible (equivalently bijective) it is an {\it isomorphism}, 
if $\nFrS_1 = \nFrS_2$ it is an                                          {\it endomorphism}, 
and, if both apply,    it is an                                          {\it automorphism}.} 
%
$Aut(\FA)$ of absolute space $\FA$ is then an obvious possibility for $\FrG$. 
However some subgroup \cite{Kobayashi} of $Aut(\FA)$ might also be desirable, 
not least because the inclusion of some such automorphisms depends on which level of mathematical structure $\sigma$ is to be taken to be physical.
I.e. $\FrG \leq Aut(\langle\FA, \sigma\rangle)$ is a more general possibility.
Such subgroups also comply with A) and stand a good chance of suitably satisfying criteria B) and C).
See Sec \ref{Mech} for examples.

\subsection{Direct implementation of Configurational Relationalism}

Suppose we are in possession of a $\FrQ$, $\FrG$ candidate pair.
Then seek to represent the generators of $\FrG$ in terms of $Q^{\sfA}, \frac{\pa}{\pa Q_{\tfA}}$ which manifestly acts on $\FrQ$,   
We can then check whether some candidate objects $O(Q^{\sfA})$ are $\FrG$-invariants by explicitly checking that these are indeed preserved by $\FrG$'s generators.  
For particles configurations, invariants are plentiful and intuitively clear for a wide range of $\FrG$, 
as a direct follow-on of the forms taken by the corresponding $\FrG$-invariants invariants in $\mathbb{R}^d$.  
E.g. for $\FrG = Eucl(d)$, separations between particles and angles between particles are of this nature; Fig \ref{More-G-Invariants} tabulates such for further $\FrG$.

\subsection{Indirect implementation of Configurational Relationalism: `$\FrG$-act $\FrG$-all' method}\label{Gact-Gall}

A given set of objects can also be interesting in a wider range of ways than being $\FrG$-invariants. 
Moreover, in some cases, invariants are unknown or nonexistent.  
There are more general concepts of `good $\FrG$ objects, such as $\FrG$-tensors (of which $\FrG$-invariants are but one example: $\FrG$-scalars)
or objects which are one of the preceding modulo a linear function of the generators.  
Here one is representing auxiliaries in terms of $q^{\sfE}$ and tangent bundle auxiliary quantities $\d g_{\sfG}$.  

\mbox{ } 

\ni Configurational Relationalism's broad strategy is the `$\FrG$-{\it act} $\FrG$-{\it all method} \cite{ARel}. 
Consider here consider some object $O$ belonging to some space of objects $\FrO$.
The $O$ may be composites of some kind of more basic variables $b$ ($Q^{\sfA}$ alone in this Sec's $\FrQ$-only setting, though further Secs such as Sec \ref{Mech} extend this).  
Such composites indeed cover far more than just $\FS$: also e.g. notions of distance, information, correlation, 
and also quantum operators and quantum versions of all of the preceding.  
In whichever case, start by applying $\FrG$-act; 
this can initially be conceived of as $\FrO \stackrel{\sFG \times}{\longrightarrow} \FrG \times \FrO$, $O \mapsto \stackrel{\rightarrow}{\FrG}_g O$.  
End by applying $\FrG$-all: some operation $\mbox{\Large S}_{g \in \sFG}$ is applied, which makes use of all of the $g^{\sfG} \in \FrG$.
This has the effect of cancelling out $\FrG$-act's use of $g^{\sfG}$, so overall a $\FrG$-invariant version of each $O$ is produced, which I denote by 
\beq
O_{\sFG-\si\sn\sv} := \mbox{\Large S}_{g \in \sFG} \, \circ \stackrel{\rightarrow}{\FrG}_g O \mbox{ } .  
\eeq
Examples of $\mbox{\Large S}_{g \in \sFG}$ include summing, integrating, averaging (group averaging is an important basic technique in Group and Representation Theory), 
taking infs or sups, and extremizing, in each case indeed meaning over $\FrG$.  
For the first two examples,
\beq
\mbox{\Large S}_{g \in \sFG} \mbox{ } \mbox{ include } \mbox{ } \sumG     \mbox{ } ,  \mbox{ }   \int_{g \in \sFG}\mathbb{D} g    \mbox{ } .                                                                            
\eeq
Barbour's Best Matching's own $\mbox{\Large S}_{g \in \sFG}$ is extremization over $\FrG$ (see Secs \ref{Comparer}--\ref{Mech}).

Finally, `Maps' can additionally be inserted between $\FrG$-act and $\FrG$-all to produce an even more general 
\beq
O_{\sFG-\si\sn\sv} := \mbox{\Large S}_{g \in \sFG} \circ \mbox{Maps } \circ \stackrel{\rightarrow}{\FrG} O \mbox{ } . 
\eeq
`Maps' covers a very general assortment of maps, though these are to all be $\FrG$-invariant; if not, $\FrG$ would act on a new type of object $O^{\prime} = \mbox{Maps} \circ O$.

\subsection{On the variety of notions of point particle configuration}\label{+RPMs}

For the most commonly considered case of $\FA = \mathbb{R}^d$, Appendices \ref{Flat-Geom} and \ref{Lie} provide many suitable $\FrG$'s; see also Fig \ref{Tab-1}.   
What physical considerations enter these choices? 
A case of note is whether scale is to be physically meaningless.  
Moreover, in directly modelling nature, disregarding scale jeopardizes standard cosmological theory without providing a viable replacement \cite{ABFO}. 
Additionally, retaining scale may enable time provision \cite{ACos2}.  
On the other hand, the metric Shape Theory is both mathematically simpler and recurs as a subproblem within the metric Scale-and-Shape theory.
Moreover, it is quite commonplace to consider physical theories with scale that possess a distinct scale-invariant phase in an `unbroken' higher-energy regime, 
by which not matching everyday experience is not necessarily the end to a theory's relevance.  

{            \begin{figure}[ht]
\centering
\includegraphics[width=0.8\textwidth]{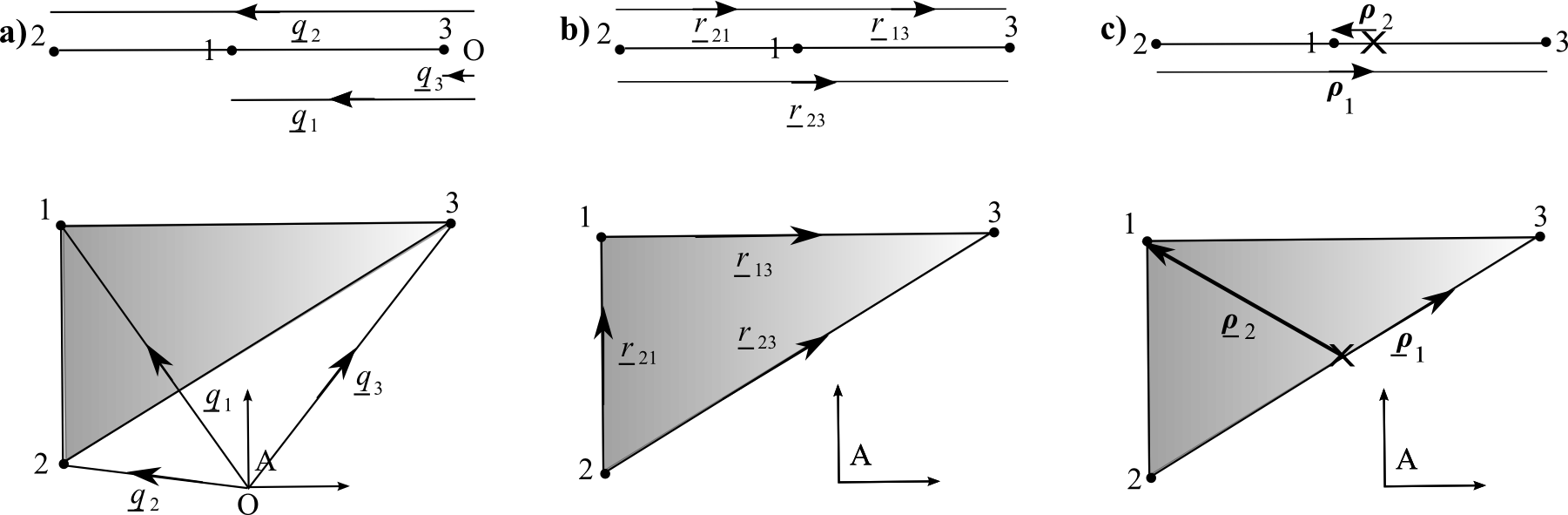}
\caption[Text der im Bilderverzeichnis auftaucht]{        \footnotesize{Coordinate systems for 3 particles
(underline denotes spatial vectors and bold font denotes configuration space quantities).
\ni a) and b) Absolute particle position coordinates ($\q_1$, $\q_2$, $\q_3$) in 1- and 2-$d$.  
These are reckoned with respect to fixed axes A and a fixed origin O.
\ni c) and d) Relative inter-particle (Lagrange) coordinates $\{\r^{IJ}, I > J\}$.
Their relation to the $\q^I$ are obvious: $\r^{IJ} := \q^J - \q^{I}$.  
In the case of 3 particles, any 2 of these form a basis; I use upper-case Latin indices $A, B, C$ for a basis of relative separation labels 1 to $n$.  
No fixed origin enters their definition, but they are in no way freed from fixed coordinate axes A.
\ni e) and f) It is more convenient to work with relative Jacobi coordinates because these diagonalize the 
`mass matrix' or kinetic metric that the moment of inertia, kinetic term and kinetic arc element are built from.  
Relative Jacobi coordinates attain this through in general being relative separations of {\sl particle clusters}. 
$\mbox{\large $\times$}$ denotes the centre of mass of particles 2 and 3. 
Convenience furthermore dictates that I take the {\sl mass-weighted} relative Jacobi coordinates $\underline{\rho}^A$.
Then the kinetic metric is just an identity array with components $\delta_{ij}\delta_{AB}$.  
Jacobi coordinates are indeed widely used in Celestial Mechanics \cite{Marchal} and Molecular Physics \cite{LR97}.} }
\label{Relative-Coordinates}\end{figure}            }

{            \begin{figure}[ht]
\centering
\includegraphics[width=0.85\textwidth]{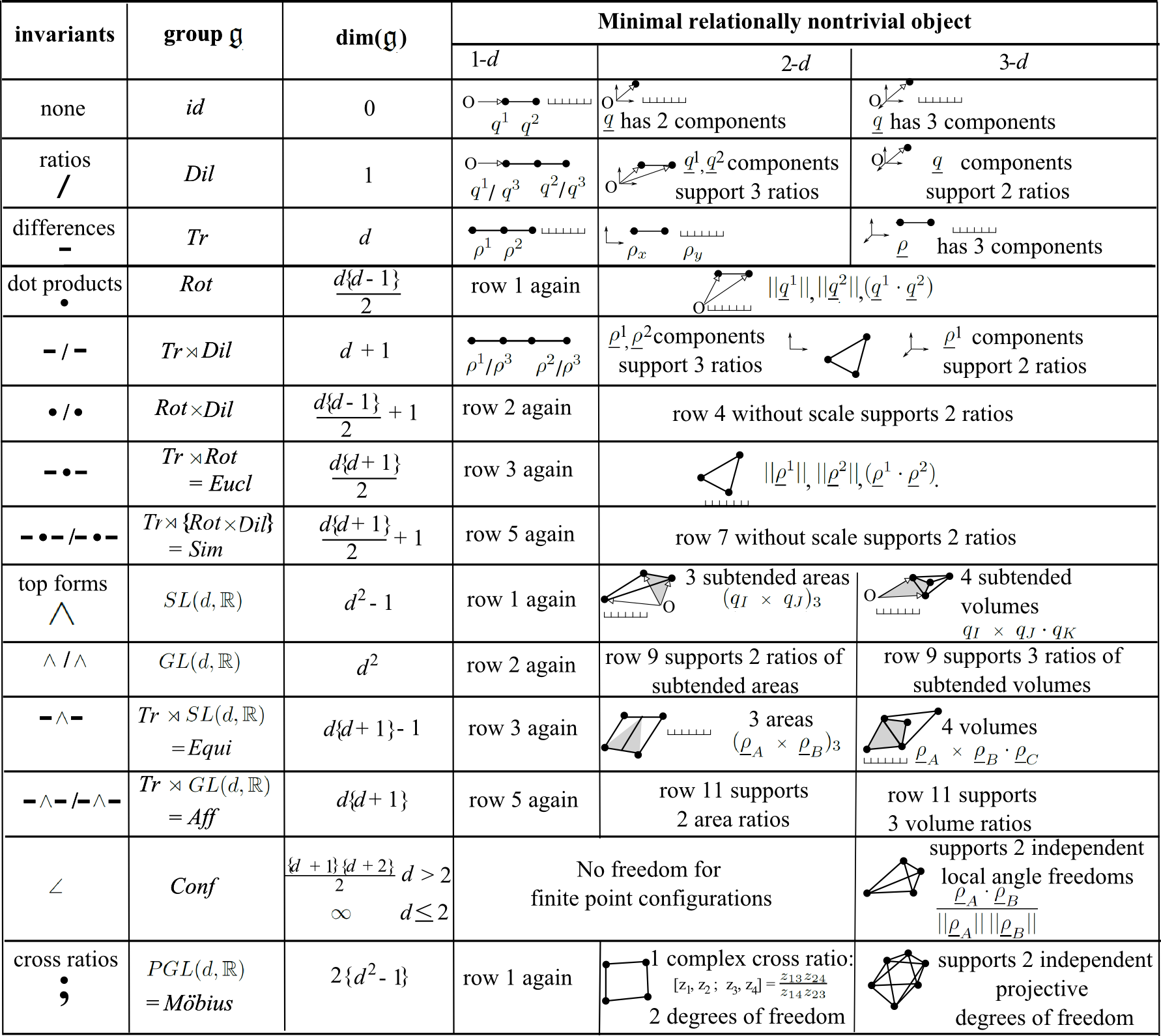}
\caption[Text der im Bilderverzeichnis auftaucht]{        \footnotesize{\ni Invariant corresponding to each group, dimension of the group and 
the minimal relationally nontrivial unit in spatial dimensions 1, 2 and 3.  
O is an absolute origin; the axis and ruler logos denote absolute orientation and absolute scale respectively.
In particular, note that the bottom six rows are interesting and new. 
In this way, the current paper covers enough new material, with a broad enough range of new applications for further researchers to work on, 
to justify introducing this useful shorthand notation for the invariants/observables, 
and the namings of all the corresponding configurations and of theories of Mechanics, and of Statistics in this and the next Figure.} }
\label{Tab-1} \end{figure}          }

\ni One possibility which has hitherto not been mentioned in setting up RPMs is that preservation of inner products $\cdot$ is not the only possibility:  
an alternative to this is preservation of $\shortcr$ products (or in dimension-independent language of forms, of exterior products $\wedge$ \mbox{ }).
This corresponds to whether infinitesimal Procrustean stretches ($d$-volume top form preserving in dimension $d$) and shears are to be physically meaningless.
These distort relative angles and of ratios of relative separations respectively. 
They combine with translations and rotations to form the `equi-top-form group' $Equi(d)$, or furthermore with the dilations to form the affine group $Aff(d)$.
This represents one way of extending $Sim(d)$ through its being a subgroup within a larger group
Each of these groups corresponds to a further known type of geometry as per Appendix \ref{Lie-Ex}.

Further issues involve whether the configurations are to be mirror image identified, and whether the particles are to be distinguishable.
Both of these issues translate to the form taken by the configuration space topology \cite{FileR}.
This involves using not necessarily $\FrQ$ but $\FrQ = \sumIN\FA/\FrG^{\prime}$ more generally,\footnote{I use a Gothic font for spaces so as to not confuse 
configurations with the configuration spaces they belong to.
Also $\mathbb{Z}_a$ is here the cyclic group of order $a$.}
in particular for discrete group $\FrG^{\prime} = \mathbb{Z}_2, \mathbb{Z}_N, \mathbb{Z}_2 \times \mathbb{Z}_N$ though partial indistinguishability is also possible.

A distinct further possibility which has not yet been mentioned in setting up RPMs stems from whether to allow the `inversion in the sphere' transformation (\ref{Inv}), 
which {\sl also} preserves angles. 
If so, special conformal transformations exist, providing a distinct extension of $Sim(d)$.
Note furthermore that the affine and conformal extensions are incompatible with each other as per (\ref{K-S}), so these two extensions cannot furthermore be composed. 
They correspond to two different `apex groups' within each of which the $Sim(d)$ hitherto used in RPMs and Shape Statistics sits as a subgroup.

This is a good point at which to note that 1-$d$ is too simple to support distinctions between a number of types of geometry. 
As well as having no continuous rotations (and its only discrete rotation coinciding with inversion), 1-$d$ has no nontrivial volume forms and so is bereft of an affine extension. 
Thus rotations and the further possibility of extension to affine transformations require dimension $\geq 2$. 
On the other hand, 2-$d$ has an infinite-$d$ conformal group. 
In fact 1-$d$ does too, though that one is less interesting through coinciding with the reparametrizations.  
See Appendix B for an outline of both of these workings). 
These invalidate 1 and 2-$d$ configurations of a finite number of particles from having a nontrivial relational theory by the counting argument A) of Sec 2.  
Thus dimension $\geq 3$ is required to investigate this possibility, though e.g. dimension 2 also provides finite subgroups of the conformal group within which $Sim(2)$ sits as a subgroup. 
E.g. the M\"{o}bius group considered in Sec \ref{Complex}; this group in turn corresponds to a type of projective geometry.\footnote{Further real projective 
variants additionally exhibit point-to-line duality, which further unusual property heralds departure from Relational {\sl Particle} Mechanics.  
Moreover, the $\mathbb{C}$ case of projective geometry additionally does not discern between lines and circles.}
 
This is also a good point at which to discuss the Relationalism of the above two extensions.  
The conformal extension's involvement of inversion amounts to replacing $\FA(d) = \mathbb{R}^d$ by $\mathbb{R}^d \cup \infty$; 
this is a new consideration in the context of RPMs and Background Independence more generally.   
On the one hand, this is adding an extra structure which might be interpreted as absolute: Riemann's notion of the `point at infinity'. 
On the other hand, appending this one point allows for a rather more general class of angle-preserving transformations to be well-defined.
The affine extension, on the other hand, remains within the usual $\FA(d) = \mathbb{R}^d$.
It amounts to modelling situations in which configurations have no {\sl overall} meaning of either relative angle (by equivalence under global shears), 
                                                                                             or of relative ratio (by equivalence under global Procrustean stretches).
Considering these in the context of RPMs is also new to the current paper.  
See Sec \ref{Mech} for discussion of issues of direct realization, and of applications for which that is not a consideration.

\mbox{ } 

\ni In the by now well studied relational space and (metric) shape space cases, in both 2- and 3-$d$ the minimal relationally nontrivial unit is the triangle. 
Barbour's well-known demonstrations of Best Matching with wooden triangles, and Kendall's method of sampling in threes leading to his spherical blackboard methodology 
-- Fig \ref{Sphe-Chop}.a -- follow. 
{\it The minimal relationally nontrivial unit is all of that type of Relationalism's smallest whole universe relational model, smallest relationally nontrivial subsystem, 
and smallest relationally nontrivial sampling unit for Shape Statistics.} 
These are ways in which minimal relationally nontrivial units are important.  
Thus in the last three columns of Fig \ref{Tab-1}, I depict and explain the form these take, 
for each of this paper's suite of subgroups of the affine group, the full conformal group in 3-$d$ and the M\"{o}bius group in 2-$d$, 
In perusing this Figure, 
it may be useful to bear in mind how the simplest affine and projective geometry theorems also require use of more points than the simplest Euclidean geometry ones \cite{Coxeter}. 

\mbox{ } 

\ni More generally, yet further models of absolute space $\FA$ might be considered, such as $\mathbb{S}^d$ or $\mathbb{T}^d$.  
E.g. for GR, both $\mathbb{S}^3$ and $\mathbb{T}^3$ have been substantially studied as compact models for space; 
$\mathbb{T}^d$ is of course also a compactification of $\mathbb{R}^d$; 
Cosmology also makes use of hyperbolic space $\mathbb{H}^3$, which admits compactifications of its own.  
RPMs based on such are then closer to GR than RPMs based on flat space.
RPMs based on $\mathbb{S}^d$ in place of $\mathbb{R}^d$ in the role of absolute space have started to be considered elsewhere \cite{FileR, ASphe}.
Not that $\mathbb{S}^2$ admits the additional interpretation and realization as an {\it observed space} model: the sky, due to which Shape Theory and Shape Statistics for it 
had already previously appeared in the literature \cite{Kendall89, Kendall}.  
The current paper does not further explore this paragraph's additional possibilities, 
nor have RPM on $\mathbb{T}^d$, $\mathbb{H}^d$ or compactifications thereof started to be investigated to date.  
None the less, it is clear that the current paper's systematics concerning invariants, groups, families of subgroups, minimal relationally nontrivial units, configuration spaces, 
configuration comparers, and constructions of Mechanics and Statistics upon (generalized) shape spaces, furthermore carries over to all these other cases as well.  
In particular, the current paper considers the different levels of geometry on open infinite flat space,  
but if one passes to whichever curved space instead, one can again contemplate a comparable range of geometrical structures thereupon.]

\section{Configuration space geometry}\label{Q-Geom}

In setting up a large number of new theories of Shape Mechanics, with underlying Shape Theories and configuration spaces, many of which are new too, if is rather 
necessary to create names and notations for the configuration spaces in question (Figure \ref{Tab-2}).  
These are conceptually important entities, whose precise mathematical nature shall one day be known in detail, 
much as \cite{Kendall, FileR} lay this out in the case of redundant similarity group. 

{            \begin{figure}[ht]
\centering
\includegraphics[width=0.9\textwidth]{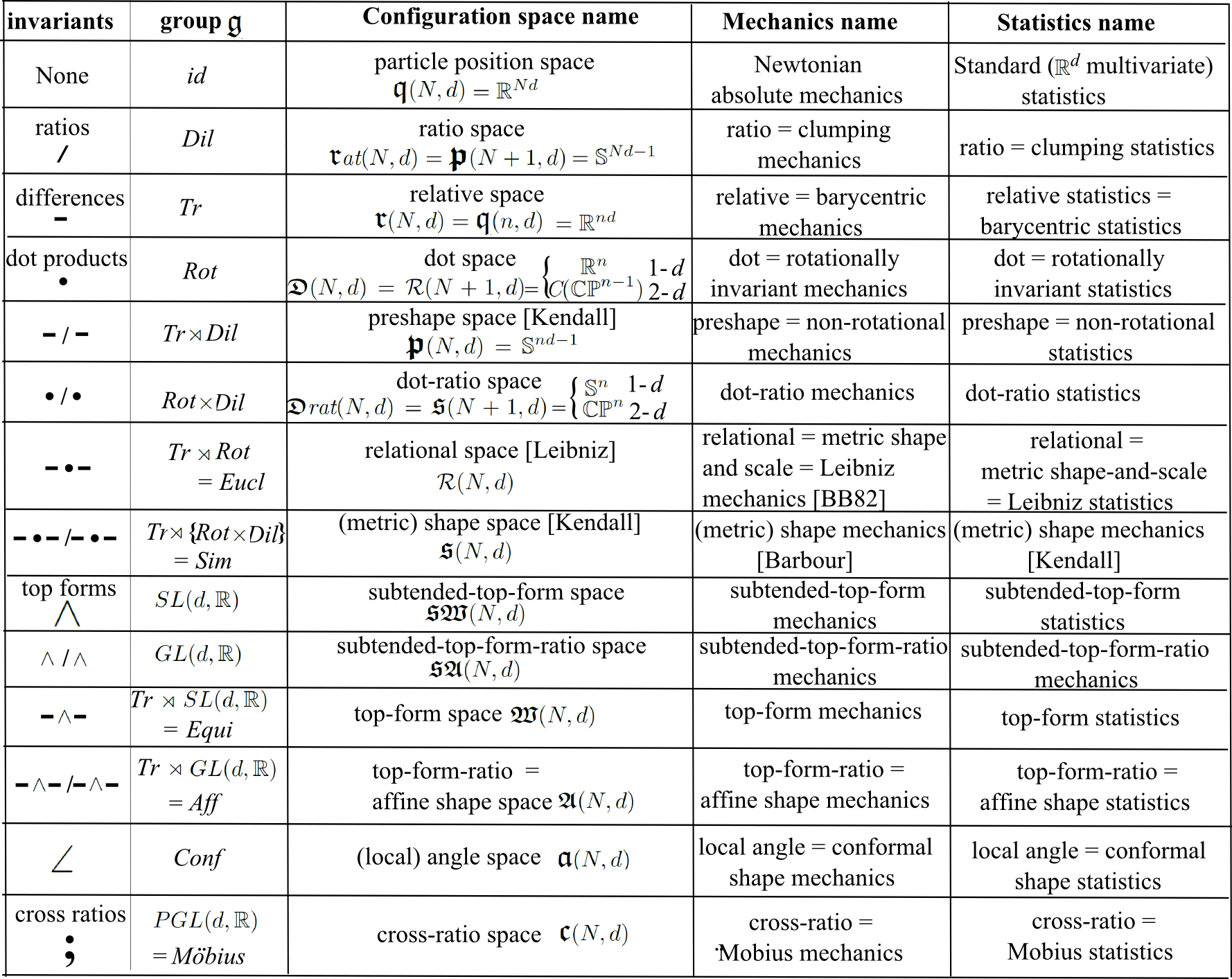}
\caption[Text der im Bilderverzeichnis auftaucht]{        \footnotesize{For each group--invariant pair, 
we give name and notation for the corresponding configuration space, and the names for the corresponding Mechanics and Statistics.
Note the two alternative namings: by group and by the theory's invariant objects. 
Note many configuration space geometry notions reduce to others, and that ones of known geometry are indicated; see \cite{Kendall, FileR, AConfig} for more about these.
N.B. that the last 6 rows are both new and of particular significance.} }
\label{Tab-2} \end{figure}          }

{\it Relative space} $\Fr(N, d) = \FrQ(N, d)/Tr(d) = \mathbb{R}^{n d}$ for $n := N - 1$.
Bases of relative inter-particle separation vectors -- Lagrange coordinates -- and of cluster separation vectors -- relative Jacobi coordinates -- are then natural thereupon 
(Fig \ref{Relative-Coordinates}.c)--f).  
In an absolute worldview, these correspond to passing to centre of mass frame, whereas in a relational worldview they correspond to absolute absolute origin being meaningless.

\mbox{ } 

\ni {\bf Useful Lemma (Jacobi pairs)}.
Within the subgroups of the affine group, the number of relational configuration spaces requiring independent study is halved, 
since each version with translations is the same as the version without with one particle more.  

\ni{\underline{Proof}. For these groups taking out the centre of mass is always equally trivial. 
Moreover, the diagonal form in Jacobi's relative $\underline{\rho}^A$ is identical in every respect with that of the mass-weighted point particles 
                             bar there being one $\underline{\rho}^A$ less. $\Box$ 

\mbox{ } 

\ni Take the form of the invariants of a geometry [Appendix A]  and apply to whichever of\foo{Bold font here denotes configuration space objects, with indices $iI$ and $iA$ 
respectively.} 
$\bq$ and $\brho$ are rendered appropriate by the assumption of material point particles [displayed in column 1 of Figs \ref{Tab-1}--\ref{Tab-2} and further laid out in Fig \ref{More-G}]. 
These are what then firstly serve as potential functional dependence, and subsequently end up being wavefunction dependencies. 
Figs \ref{Tab-2} and \ref{More-G-Invariants} then name the corresponding configuration spaces. 

\mbox{ } 

\ni If absolute axes are also to have no meaning, the remaining configuration space is 
\beq
\mbox{\it relational space }  \mbox{ } \bigr(N, d) := \FrQ(N, d)/\mbox{$Eucl$($d$)} \mbox{ } ,    
\eeq
of dimension $nd$ -- $d\{d - 1\}/2$ = d\{2$n$ + 1 -- $d$\}/2: in particular, $N - 1$ in 2-$d$, $2N - 3$ in 2-$d$ and $3N - 6$ in 3-$d$.
If, instead, absolute scale is also to have no meaning, the configuration space is Kendall's {\it preshape space} \cite{Kendall} $\FP(N, d) := \Fr(N, d)/\mbox{$Dil$}$, 
of dimension $nd - 1$.  
If both absolute axes and absolute scale are to have no meaning, then the configuration space is Kendall's \cite{Kendall} 
\beq
\mbox{\it shape space } \mbox{ } \FrS(N, d) := \FrQ(N, d)/\mbox{Sim}(d) \mbox{ } . 
\eeq
This is of dimension $Nd - \{d\{d + 1\}/2 + 1\} = d\{2n + 1 - d\}/2 - 1$; in particular $N - 2$ in 1-$d$, $2N - 4$ in 2-$d$ and $3N - 7$ in 3-$d$; 
as well as featuring in accounts of RPMs \cite{FORD, FileR}, 
it is well-known from the Shape Geometry and the Shape Statistics literatures \cite{Kendall84, Kendall89, Kendall}.
Also note that $\FP(N, 1) = \FrS(N, 1)$, since there are no rotations in 1-$d$.  
The above quotient spaces are taken to be not just sets but also normed spaces, metric spaces, topological spaces, and, where possible, Riemannian geometries.      
Their analogy with GR's configuration spaces is laid out in Fig \ref{Q-RPM-GR}. 

\mbox{ } 

\ni Note that the Jacobi pairs simplification does not apply within those further groups that include the special conformal transformations $K_i$.    
This is because of the commutation relation (\ref{K-P}), by which translations cease to be so trivially removable.  
Also contrast the conformal case's pure angle information with the similarity case's mixture of angle and ratio information.

See Fig \ref{More-G-Invariants} for an outline of further subgroups of Conf($d$), 
alongside indication of other combinations of generators which also fail to close as groups for the reasons stated.

The configuration space level can have a metric geometry (or in reduced cases generally a stratified such), 
even if the original configurations sit in a geometry with less structure than that. 
This is entirely possible because the map from space to the space of spaces need {\sl not} be category-preserving.

Finally, while two area ratios have the range of a quadrant or a whole plane if signed, it is not a priori likely for these to carry a flat metric. 

{            \begin{figure}[ht]
\centering
\includegraphics[width=0.75\textwidth]{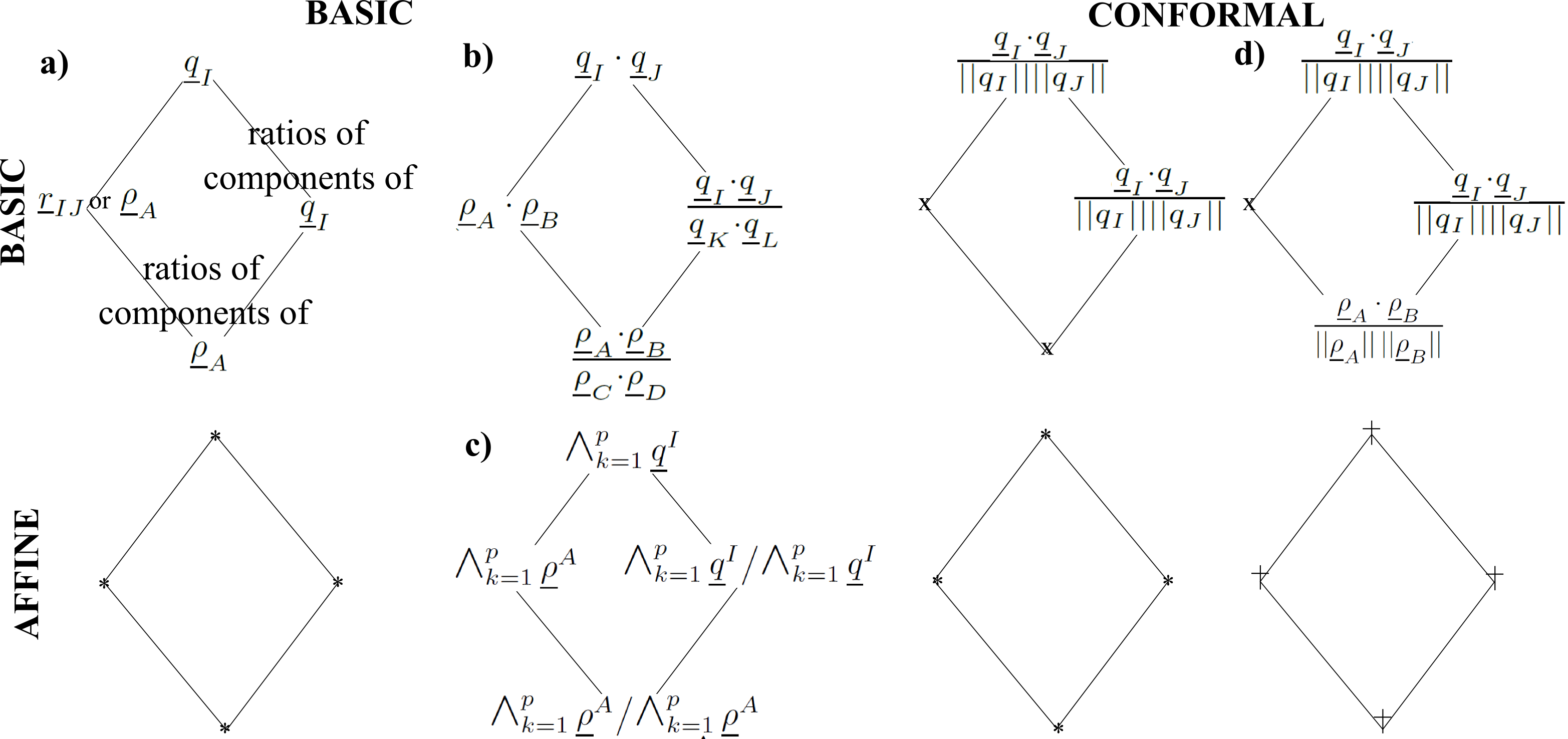}
\caption[Text der im Bilderverzeichnis auftaucht]{        \footnotesize{ \ni Layout of $\FrG$-invariant contents, 
in the pattern following on from Fig \ref{More-G-Lie-2}'s layout of which combinations of generators are group-theoretically allowed.  
Note that four of the subgroups of $Conf(d)$ have the same invariants; this is due to incorporating the special conformal transformation being rather restrictive.
} }
\label{More-G-Invariants} \end{figure}          }
%
{            \begin{figure}[ht]
\centering
\includegraphics[width=1.0\textwidth]{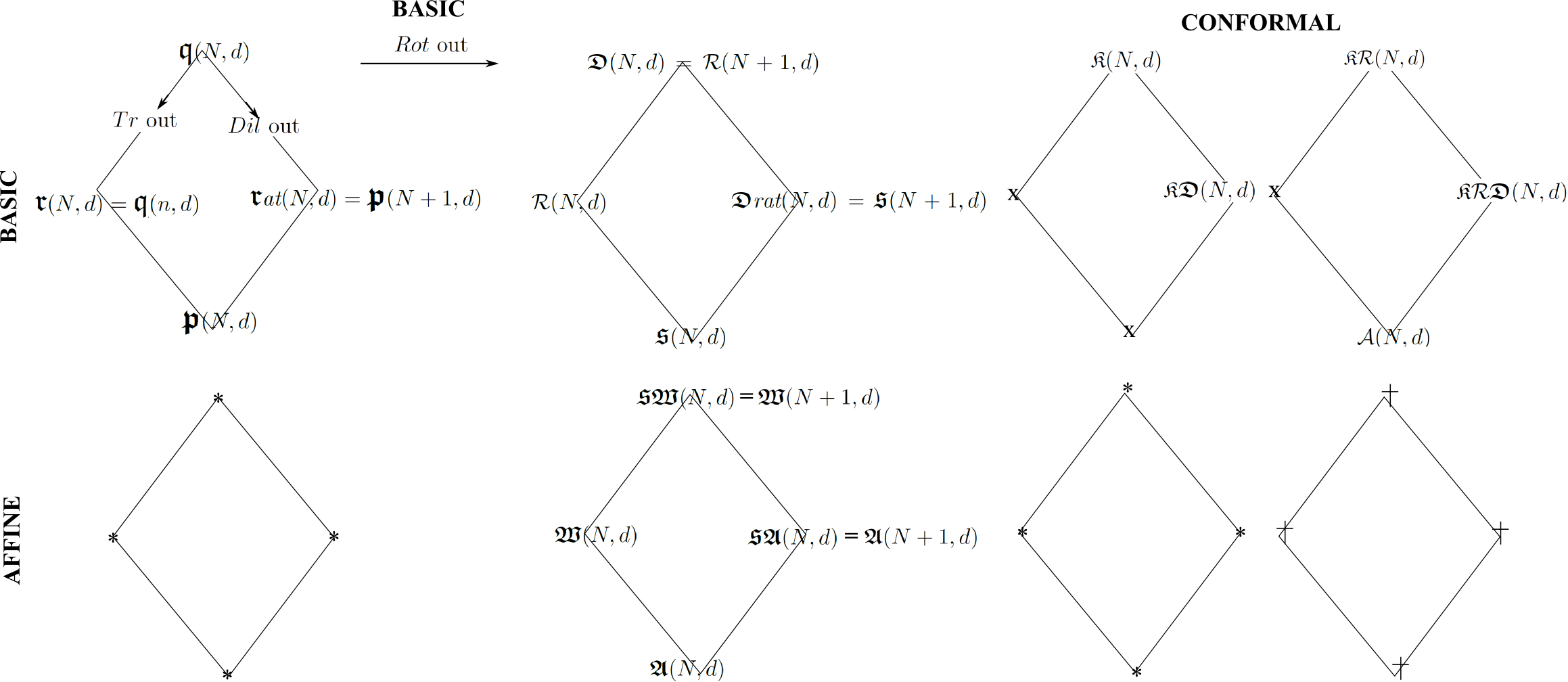}
\caption[Text der im Bilderverzeichnis auftaucht]{        \footnotesize{ \ni  Whereas Fig \ref{Tab-2} mentions the most commonly considered quotients, 
here are a number of further possibilities. 
These are once again displayed in parallel with Fig \ref{More-G-Lie-2}'s layout of which combinations of generators are group-theoretically allowed.  
Useful simplifying relations by which some of these configuration spaces are mathematically very similar to others are also indicated, 
such as the Jacobi parings and which arise from quotienting out subgroups of whichever of the affine and conformal groups.} }
\label{More-G} \end{figure}          }
%
{            \begin{figure}[ht]
\centering
\includegraphics[width=0.6\textwidth]{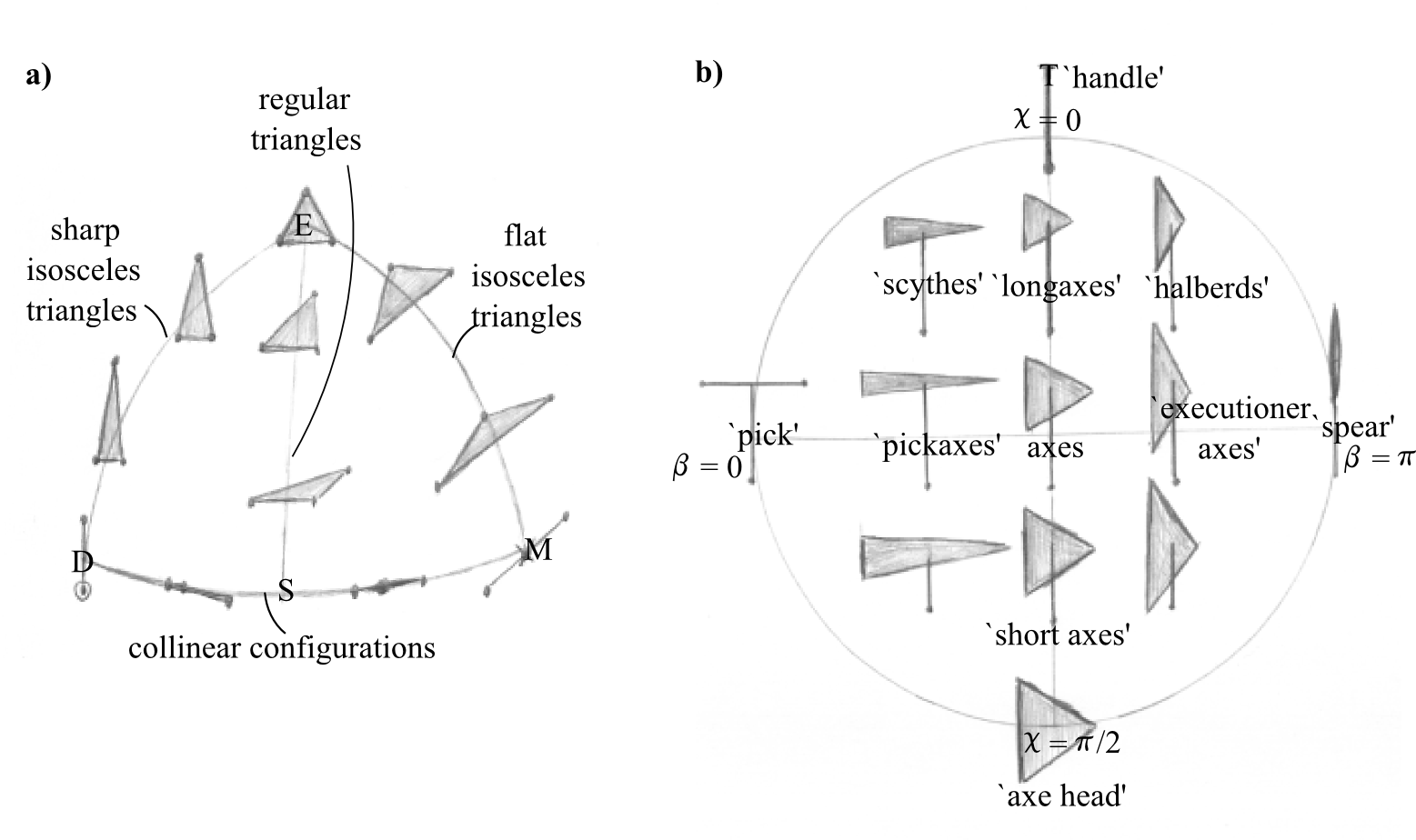}
\caption[Text der im Bilderverzeichnis auftaucht]{        \footnotesize{a) Unlabelled triangles' {\it spherical blackboard} 
(or half of it; triangleland is either $\mathbb{S}^2$ or some regular portion thereof, depending on exactly how the configurations are being modelled \cite{AConfig}).    
Here `regular' means that the base and median partial moments of inertia are equal, D is a double collision and M is a merger (third particle at the centre of mass of the other two).
If the triangles are labelled and with mirror images distinct, the whole sphere is realized.
b) {\it Complex projective chopping board} of axe configuration.
This figure concentrates on the two ratio coordinates $\beta$ and $\chi$, suppressing the further variety in relative angle coordinates $\phi$ and $\psi$ 
($\beta, \chi, \phi\ \psi$ are Gibbons--Pope type coordinates, see e.g. \cite{QuadI}).} }
\label{Sphe-Chop} \end{figure}          }

\mbox{ }  

\ni Considering larger units than the minimal one is valuable not only since furtherly relational theories need some such, 
but also because some applications need more system complexity.  
It is insightful here to point out that Montgomery's falling cat\footnote{This involves 2-rod configurations, corresponding to configuration space $\mathbb{RP}^2$.} 
and the relational triangle are {\sl fully minimal} robotic models. 
E.g. envisage the space of triangles not as a bunch of rigid wooden shapes but as the shapes that can be formed by a flexible, extendible entity. 
This perspective involves {\sl paths in configuration space}. 
Models along the above lines rather quickly acquire complexity.
Indeed \cite{QuadI, QuadIII} considered the K-shaped clustering of three relative Jacobi coordinate vectors presentation of quadrilaterals as axe configurations. 
The underlying shape space in this case is $\mathbb{CP}^2$: Fig \ref{Sphe-Chop}.b), and the relational space is $C(\mathbb{CP}^2)$.
I now point out that axes are already complex enough tools to have very different applications according to angles and proportions. 
Thus a `robotic axe' reinterpretation of the model shape space of axes is {\sl already} a model of a {\sl significantly adaptable} robotic tool.
Finally robotic models may also eventually expected to enter foundational Theoretical Physics, along the lines of Hartle's IGUS \cite{IGUS}.
I.e the {\it Information Gathering and Utilizing System} model concept could well receive a classical and then quantum robotic implementation.

\section{Configuration comparers}\label{Comparer}

{            \begin{figure}[ht]
\centering
\includegraphics[width=1.0\textwidth]{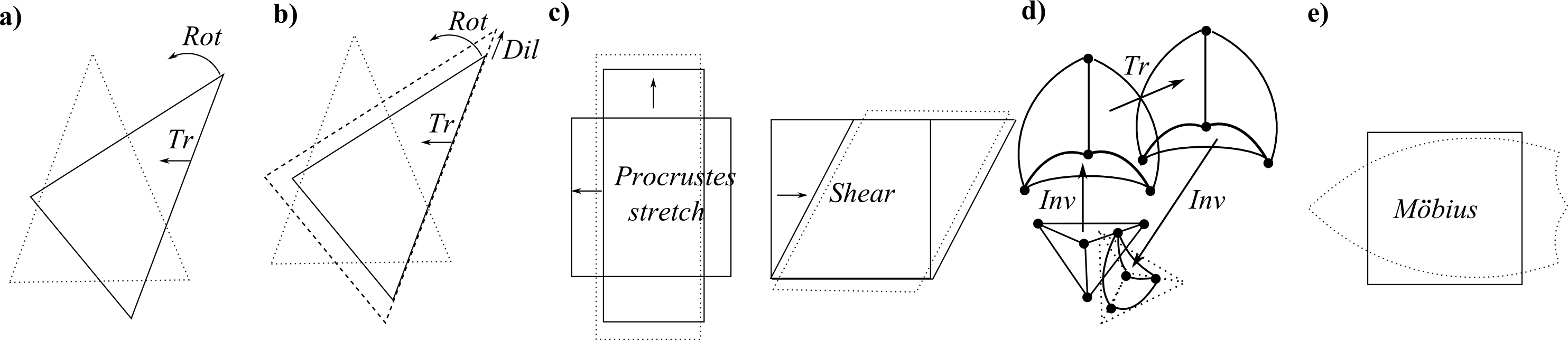}
\caption[Text der im Bilderverzeichnis auftaucht]{        \footnotesize{5 notions of matching shapes keep the dotted one fixed and perform transformations.
a) Barbour's wooden triangles. 
b) The same, now for overhead projector slides. 
Affine matching adds c) to this and conformal matching adds d) 
e) is  M\"{o}bius matching.} }
\label{BM-Suite} \end{figure}          }

Various such can be built from $\FrQ$'s kinetic metric's\footnote{This exists independently of whether it is contracted into velocities or changes; 
e.g. moment of inertia is this metric contracted into mechanical configurations themselves.
It only provides a norm if it is positive-definite.}
$\mbox{\boldmath$M$}$ inner product and norm \cite{Kendall84, Kendall, BB82, DeWitt70, FileR}   
\be
\mbox{(Kendall Dist)} = (\mbox{\boldmath$Q$}, \mbox{\boldmath$Q$})\mbox{}_{\mbox{\scriptsize\boldmath$M$}} \mbox{ } ,    
\label{ProtoKenComp}
\ee
\be
\mbox{(Barbour Dist)} = ||\d \mbox{\boldmath$Q$}||_{\mbox{\scriptsize\boldmath$M$}}\mbox{}^2 \mbox{ } , 
\label{ProtoBarComp}
\ee
\be
\mbox{(DeWitt Dist)} = (\d \mbox{\boldmath$Q$}, \d \mbox{\boldmath$Q$}^{\prime})_{\mbox{\scriptsize\boldmath$M$}} \mbox{ } .  
\label{ProtoDeWittComp}
\ee

\mbox{ } \mbox{ } Next, if there is additionally a physically irrelevant $\FrG$ acting upon $\FrQ$,  

\be
\mbox{(Kendall $\FrG$-Dist)} = (\mbox{\boldmath$Q$} \cdot \stackrel{\longrightarrow}{\FrG_{g}} \mbox{\boldmath$Q$}^{\prime})_{\mbox{\scriptsize\boldmath$M$}} \mbox{ } ,  
\label{Kend}
\ee 
\be
\mbox{(Barbour $\FrG$-Dist) } =  ||\d_{\sg}\mbox{\boldmath$Q$}||_{\mbox{\scriptsize\boldmath$M$}}\mbox{}^2 \mbox{ } \mbox{ and }  
\label{Barb}
\ee
\be
\mbox{(DeWitt $\FrG$-Dist)} = 
(\stackrel{\longrightarrow}{\FrG}_{\d g} \mbox{\boldmath$Q$}, \stackrel{\longrightarrow}{\FrG}_{\d g} \mbox{\boldmath$Q$}^{\prime})_{\mbox{\scriptsize\boldmath$M$}} 
\mbox{ } .		
\label{DeWi}
\ee
Then $\FrG$-all moves -- such as integral, sum, average, inf, sup or extremum -- can be applied, after insertion of Maps if necessary. 
This is the first publication to consider this three-way comparison, DeWitt's own approach being well-known from the foundations of GR as a dynamical system \cite{DeWitt70}.

Then e.g. (\ref{Barb}) subjected to the $\times \sqrt{2W}$ and integration maps before a $\FrG$-all extremum move gives Best Matching; 
this can furthermore now be recognized as a subcase of a weighted path metric.
(\ref{Kend}) itself differs from the other two cases in using a {\sl finite} group action to the other two cases' infinitesimal ones.
In another sense, it is (\ref{Kend}) and (\ref{DeWi}) which are akin: compare two distinct inputs versus (\ref{Barb}) working around a single input.  
`Comparers' then have a further issue: if $\mbox{\boldmath$M$} = \mbox{\boldmath$M$}(\fQ)$, does one use $Q_1$ or $Q_2$ in evaluating $\mbox{\boldmath$M$}$ itself? 
This situation does not arise in the $\mathbb{R}^n$ shapes context of Kendall,
but it does in DeWitt's GR context; he resolved it in the symmetric manner, i.e. using $Q_1$ and $Q_2$ to equal extents.  

\mbox{ } 

\ni Furthermore, {\sl how good} the `best fit' is can be assessed in substantially geometrically general cases by making set of relational objects out of primed and unprimed vertices 
(Fig \ref{Matching-2}.a), to which the corresponding notion of Shape Statistics is to be applied.  
In the present case, these are triangles that one can test against the $\epsilon$-bluntness criterion. 
%
{            \begin{figure}[ht]
\centering
\includegraphics[width=0.4\textwidth]{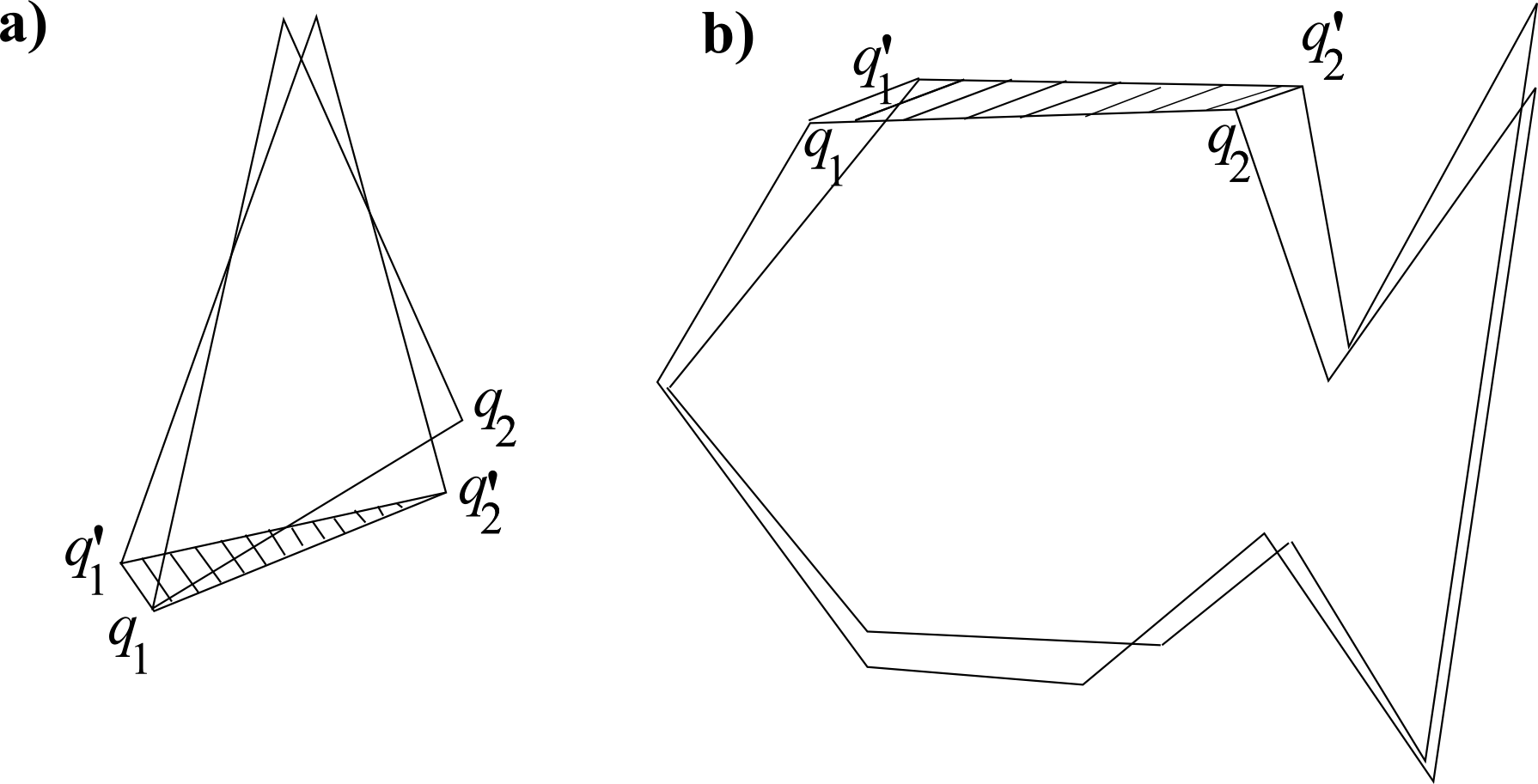}
\caption[Text der im Bilderverzeichnis auftaucht]{        \footnotesize{a) Assessing matched triangles.
b) Assessing matched fish (2-$d$ image version), 
corresponding to Thompson's observation that there are two species of fish whose shapes are approximately related by a scale and shear transformation \cite{Thompson}.
Since matching has already occurred, the images being compared can be taken to involve equal-area polygons, thus involving equiareal, rather than full affine, mathematics.
In each case, consider the minimum relationally nontrivial figures formed between the two matched figures.  
E.g. are these a) significantly blunt, b) significantly lacking in area?
} }
\label{Matching-2} \end{figure}          }

As a first new example, consider affine space.  
Its $d$-volume top form construct is not amenable to a notion of distance out of being antisymmetric and thus not obeying symmetry, separation, or, if signed, positivity.
Affine space can however be equipped as a metric space, giving a $\mbox{Dist}(\bq, \bG\,\bq)$ style finite comparer \cite{Fitzgibbon} for $\bG \, \in \, GL(d, \mathbb{R})$.  
For instance this is a useful tool in image recognition, which is affine to leading order \cite{Fitzgibbon}. 
This corresponds to how images can be stretched and sheared as well as enlarged, translated and rotated, depending on the exact modelling assumptions made.
One application of this is to nearby objects, for which many viewpoints can be attained `by walking to the other side' of the object.  
One interesting proposition then is whether two pictures feature the same object viewed from different angles.
A second application is that the former proposition also arises within hypothetical smaller closed-topology universes \cite{Spots2}, 
and within the setting of multiple images from gravitational lensing \cite{Ehlers}.
However, the first of these has additional evolution effects (the multiple images would generally correspond to the object's configuration in different aeons). 
The second has additional distortion effects (the geometry around a gravitational lens is {\sl not} flat, taking one outside the scope of the current paper). 
A third application concerns studying populations of anisotropic astrophysical objects (most obviously galaxies). 
Then does one's collection of images feature a single such viewed from multiple angles, or is it a mixture of objects from distinct populations?  
Furthermore, are we seeing these objects at entirely random orientations, or do they exhibit a statistically significant pattern of orientations?  

\mbox{ } 

\ni As further new examples, see the next two Secs for many new instances of Best Matching.

\section{Theories of Shape Mechanics}\label{Mech}

This application is to whole-universe models, for which Temporal Relationalism applies.
I.e. there is no time at the primary level for the universe as a whole.
This is best implemented by a geometrical action, which is mathematically dual to an action making no use of parametrization, 
which upon introducing a parametrization is then invariant under reparametrization. 
At least one primary constraint must then follow from any reparametrization-invariant action due to a well-known argument of Dirac's \cite{Dirac}.  
In the case of Mechanics, this gives an equation which is usually interpreted as an energy conservation equation. 
However in the present context this is to be reinterpreted as an {\it equation of time} \cite{B94I}. 
Indeed rearranging it gives an expression for emergent Machian time:  a concrete realization resolution of primary-level timelessness by Mach's `time is to be abstracted from change'.
In the case of metric Scale-and-Shape Mechanics, this emergent time amounts to a relational recovery of a quantity which is more usually regarded as Newtonian time.
Note that more generally handling Mechanics models in a temporally relational manner requires a modified version of the Principles of Dynamics as laid out in \cite{TRiPoD}.
The particular cases of geometrical actions in the next two sections are all \it Jacobi-type} action \cite{Lanczos}, corresponding to a Riemannian notion of geometry.

The case of Relationalism most usually considered as an RPM is metric Scale-and-Shape Mechanics \cite{BB82}, 
as motivated by the direct approximate physical regime in which Mechanics is used.
Note however that Sec \ref{Comparer}'s example of the affine transformations arising in the study of images 
shows how the `direct physical regime's geometry' may not be the only one of relevance.   
Finally, if fundamental theory is under consideration, the defects of Newtonian theory itself may well not be the only useful guideline.
E.g. scale-invariant and conformal models are often considered in High-Energy Physics (c.f. Sec 2.3).  
Also RPM has been argued to be of substantial value \cite{Kuchar92, EOT, KieferBook, FileR, AConfig} as a model arena for GR's own dynamical structure, configuration spaces, 
and of some of GR's Background Independence aspects and corresponding Problem of Time facets.
Viewed in this way, the mathematically simpler 1- and 2-$d$ RPMs -- which already exhibit many of the features of GR that are emulated by RPMs -- 
are often preferable to the 3-$d$ RPMs, many of whose extra complexities are not aligned with GR's extra complexities. 
Moreover, this GR model arena perspective also leaves a number of other modelling assumptions for RPMs open.
See Appendix C in this regard, both for metric shape RPM and for the new RPMs presented below.

\subsection{Metric Shape-and-Scale Mechanics (alias Euclidean RPM)}

Let us begin with the most familiar case of RPM: the Barbour--Bertotti 1982 theory, albeit reformulated in terms of relative Jacobi coordinates. 
We will then show how the other RPMs arise from various sources of geometrical variety in this setting.  
The action for this RPM is \cite{FileR}\footnote{A previous such theory existed \cite{BB77-cum-prehistory} 
but was built in another manner. 
This previous theory fails to fit mass anisotropy bounds, and ceases to look natural once cast in reduced variables.  
\cite{BB82} also remains simple all the way down its reduction process \cite{FileR}, whereas \cite{BB77-cum-prehistory} does not. 
Instead there is a distinct \cite{BB77-cum-prehistory} type theory \cite{FileR} which is simple in each stage of the reduction process' variables. 
These not matching up gives a further theoretical reason to favour \cite{BB82}, 
and theories built similarly to it for other $\nFrG$, rather than  \cite{BB77-cum-prehistory} type theories.}

\be
S[\urho^A, \underline{b}] = \sqrt{2}\int \md s\sqrt{W} \mbox{ } , \mbox{ }  
\md s = ||\d_{\underline{b}} \brho||     \mbox{ }, \mbox{ } 
d_{\underline{b}} \, \urho^A := \md \urho^A - \md \underline{b} \cr \urho^A \mbox{ } ,
\label{S-MSS}
\ee
for $W := E - V$ for $E$ the total energy and $V$ the potential, which in this case is of the form $V(-\cdot-)$.   
Also the index $A$ runs from $1$ to $n := N - 1$: the number of independent inter-particle cluster vectors, where $N$ is the number of particles itself.

The quadratic constraint\footnote{In this paper, I use the calligraphic font to denote constraints.}
\beq
{\cal E} := ||\bpi||^2/2 + V(-\cdot-) = E
\label{E-MSS}
\eeq
then follows as a primary constraint.
This is often interpreted as an energy constraint in the nonrelational and subsystem contexts, 
but is to be interpreted as an {\it equation of time} in the relational whole-universe context.
Also, the 
\be
\mbox{(zero total angular momentum of the universe)} \mbox{ } , \mbox{ } \mbox{ } \underline{{\cal L}} := \sumAn \urho^A \cr \upi_A = 0 \mbox{ }   
\label{ZAM}
\ee
follows as a secondary constraint from varying with respect to $\underline{b}$; thus it results from implementing Configurational Relationalism. 

\mbox{ } 

\ni [The original version \cite{BB82} has the $\urho^A$ to $\underline{q}^I$ of the above, 
with an extra translational best matching correction $- \mbox{ } \d\underline{a}$ resulting in an extra 
\beq
(\mbox{\it zero total momentum constraint}) \mbox{ } \mbox{ } \underline{{\cal P}} \mbox{ } :=  \mbox{ } \sumIN\underline{p}_I = 0 \mbox{ } , 
\label{ZM}
\eeq
for $\underline{p}_I$ the momentum conjugate to $\underline{q}^I$.]

\subsection{Metric Shape Mechanics (alias similarity RPM)}\label{Sim}

In this case -- the theory originally due to Barbour 2003 \cite{B03} and somewhat reformulated as per \cite{FORD, FileR} --
\be
S[\urho^A, \underline{b}, c] = \sqrt{2}\int \md s\sqrt{W}                                                             \mbox{ } , \mbox{ } \mbox{ } 
\md s = ||\md_{\underline{b}, c} \brho||/\rho                                                       \mbox{ } , \mbox{ } \mbox{ }
\md_{\underline{b}, c} \urho^A := \md\urho^A - \md \underline{b} \cr \urho^A - \md c \, \urho^A     \mbox{ }  
\label{S-MS}
\ee
for $V = V(-\cdot-/-\cdot-)$ and $\rho := \sqrt{I}$ for $I$ the total moment of inertia.

This gives 
\beq
{\cal E} := I ||\bpi||^2/2 + V(-\cdot-/-\cdot-) = E
\label{E-MS}
\eeq
as a primary constraint, (\ref{ZAM}) again from variation with respect to $\underline{b}$, and, from variation with respect to $c$ \cite{B03},  
\be
\mbox{(zero total dilational momentum of the universe)} \mbox{ } , \mbox{ } \mbox{ } {\cal D} := \sumIN  \urho^A \cdot \upi_A = 0 \mbox{ } .
\label{ZDM}
\ee

\subsection{3-$d$ conformal Shape Mechanics, alias local angle Mechanics}

In fact, the further special conformal Best Matching can be appended into the $\underline{q}^I$ version of (\ref{S-MS}) to give ($I$ runs from $1$ to $N$)  
\be
S[\underline{q}^I, \underline{a}, \underline{b}, c, \underline{k}] = \sqrt{2}\int \md s\sqrt{W}                                                                                                    \mbox{ } , \mbox{ } \mbox{ } 
\md s = ||\md_{\underline{a}, \underline{b}, c, \underline{k}} \bq||/\sqrt{I}                                                     \mbox{ } , \mbox{ } \mbox{ }
\md_{\underline{a}, \underline{b}, c, \underline{k}} q^{Ia} := \d q^{Ia}   - \d a^a - ( \d \underline{b} \cr \underline{q}^I)^a 
                                                             - \d c \, q^{Ia} - \{q^{I\,2}\delta^{ab} - 2 q^{Ia} q^{Ib}\} \d k_b   
\label{S-CS}
\ee
for $V$ now of the form $V(\angle)$.  
This is a first RPM theory that is new to this paper.  
Then    variations with respect to $\underline{a}$, $\underline{b}$ and $c$ yield (\ref{ZM}), and the $\underline{q}^I$ counterparts of (\ref{ZAM}) and (\ref{ZDM}) respectively.
But now also variation with respect to $\underline{k}$ produces the 
\beq
\mbox{({\it zero total special conformal momentum constraint})} \mbox{ } \mbox{ } {\cal K}_a := \sumIN \{q^{I\,2} {\delta_a}^b - 2q^I_a q^{Ib}\}p_{Ib} = 0 \mbox{ } .
\label{ZSCM}
\eeq
The quadratic constraint arising as a primary constraint is now 
\beq
{\cal E} := I ||\bpi||^2/2 + V(\angle) = E\mbox{ } .
\label{E-CS}
\eeq
Then the linear constraints close as per the conformal algebra, and $\scK_a$ manages to commute with ${\cal E}$ also.
This set-up works similarly for $d > 3$; this just requires a different presentation for the larger amounts of $Rot(d)$.

\mbox{ } 

\ni One further motivation for this model is that, while Similarity Mechanics is already to some extent a useful model of GR's conformal superspace (Appendix C), 
Conformal Shape Mechanics is surely a better model of this.

\subsection{Affine Scale-and-Shape Mechanics in 2-$d$, alias area mechanics}

As a second theory of RPM new to this paper, suppose that one proceeds by extending Sec \ref{Sim}'s construct to include affine Best Matching, so  
\beq
S[\urho^A, b, e, f] = \sqrt{2}\int \d s\sqrt{W} \mbox{ } , \mbox{ } \mbox{ } \d s = ||\d_{SL}\brho|| 
\label{S-A-No} 
\eeq
for $\d_{SL}\urho^A := \d \urho^A - \d \underline{s} \, \underline{\underline{\underline{S}}} \, \urho^A$ the $SL(2, \mathbb{R})$ best-matched derivative, 
for $\d\underline{s}$ the 3-vector $[\d f, \d e, \d b]$ of auxiliaries and 
\beq
\underline{\underline{\underline{S}}} := \mbox{\Huge[} \mbox{\Huge(}\stackrel{\mbox{1 \,\, 0}}{\mbox{0 \, --1}}      \mbox{\Huge)},
                     \mbox{\Huge(}\stackrel{\mbox{0 \,\, 1}}{\mbox{1 \,\,\, 0}}   \mbox{\Huge)}, 
	                 \mbox{\Huge(}\stackrel{\mbox{0 \, --1}}{\mbox{1 \,\,\, 0}}    \mbox{\Huge)}\mbox{\Huge]}^{\sT} \mbox{ }  , 
\label{SL-Array}
\eeq
alongside $V = V(-\shortcr-)$ in this case.  
Then 
\be
{\cal E} := ||\bpi||^2/2 + V(-\shortcr-) = E
\label{E-A-No}
\ee 
arises as a primary constraint, and the linear {\it zero total} $SL(2, \mathbb{R})$ {\it momentum constraint}
\beq
\underline{{\cal S}} := \sumAn \urho^A \underline{\underline{\underline{S}}} \, \upi_A
\label{ZSLM}
\eeq
arises from variation with respect to $\underline{s}$.
This includes ${\cal L}$ again from $b$-variation alongside new {\it zero total Procrustean momentum} ${\cal P} \sr$ and {\it shear momentum} ${\cal S }\sh$ constraints.  
Explicitly, 
\beq
{\cal P}\sr := \rho_x\pi_x - \rho_y\pi_y \mbox{ } , \mbox{ } \mbox{ } {\cal S}\sh := \rho_x\pi_y + \rho_y\pi_x \mbox{ } .
\eeq
Then these last two linear constraints fail to Poisson-brackets close with the candidate theory's ${\cal E}$.  
Thus this constitutes an example of Best Matching failing as an implementation due to the outcome of the Dirac Algorithm \cite{Dirac}.
In such cases, one is to decide which part(s) of the triple $\FrQ$, $\FrG$, $S$ are to be modified.  
The present case has a clear solution: the $||\mbox{ }||$ structure ceases to have any business in an affine theory!
In this way, the id to Conf family of further Best Matching appendings within the normed form of kinetic line element is not a general procedure. 
What is more general is my `good $\FrG$ objects' approach (Sec \ref{Gact-Gall}), by which $||\mbox{ }||$ is recognized to be illicit at the outset for a $\FrG$ as redundant as $Equi(2)$.

According to that, take instead
\beq
\d s^2 = \sumABn (\d_{SL}\urho^A \cr \d_{SL}\urho^B)_{\mbox{}_{\mbox{\scriptsize 3}}} \mbox{  } . 
\label{S-A}
\eeq
Then e.g. focusing on the smallest relationally nontrivial case, $\sumABcycles3 (\d_{SL}\urho^A \cr \d_{SL}\urho^B)_{\mbox{}_{\mbox{\scriptsize 3}}}$
inverts nicely in the change to momentum sense, giving the primary constraint 
\beq
{\cal E} := \sumABcycles3 \upi^A \cr \upi^B/2 + V(-\shortcr-) = E \mbox{ } , 
\label{E-A}
\eeq
alongside the same linear $\underline{{\cal S}}$ as before. 
This mechanical model is indeed consistent.

\subsection{Affine Shape Mechanics in 2-$d$, alias Area Ratio Mechanics}

A third theory of RPM new to this paper has 
\ni\beq
S[\urho^A, b, c, e, f] = 
\sqrt{2}\int \d s \sqrt{W} \mbox{ } , \mbox{ } \mbox{ } \d s^2 = \sumABn \d_{GL}\urho^A \cr \d_{GL}\urho^B \mbox{\Large /} \sumCDn \urho^C \cr \urho^D \mbox{ } . 
\label{S-Aff}
\eeq 
Here, $\d_{GL} \urho^A  := \d \urho^A - \d \underline{g} \, \underline{\underline{\underline{G}}}  \, \urho^A$ the $GL(2, \mathbb{R})$ best-matched derivative, 
for   $\d \underline{g} := [\d f, \d e, \d b, \d c]$ auxiliaries and 
\beq
\underline{\underline{\underline{G}}} := \mbox{\Huge[} \mbox{\Huge(} \stackrel{\mbox{1 \,\,\, 0}}{\mbox{0 \, --1}}     \mbox{\Huge)}, 
                                         \mbox{\Huge(} \stackrel{\mbox{0 \,\, 1}}{\mbox{1 \,\,\, 0}}   \mbox{\Huge)},  
	                                     \mbox{\Huge(} \stackrel{\mbox{0 \, --1}}{\mbox{1 \,\,\, 0}}    \mbox{\Huge)}, 
		                                 \mbox{\Huge(} \stackrel{\mbox{1 \,\,\,0}}{\mbox{0 \,\,\, 1}}  \mbox{\Huge)}  \mbox{\Huge]}^{\sT} \mbox{ } .
\label{GL-Array}
\eeq		
Also $V = V(-\shortcr-/-\shortcr-)$.

This results in a the same linear constraints as in the previous Subsec plus ${\cal D}$, 
the four of which can be packaged as the {\it zero total} $GL(2, \mathbb{R})$ {\it momentum constraint}
\beq
\underline{{\cal G}} := \sumAn \urho^{A} \, \underline{\underline{\underline{G}}} \, \upi_{A}   \mbox{ } .  
\label{ZGLM}
\eeq
In the smallest relationally nontrivial case $n = 3$, using once again the sum of cycles combination, the primary constraint is 
\beq
{\cal E} := \sumABcycles3 (\urho^A \cr \urho^B)_{\mbox{}_{\mbox{\scriptsize 3}}} \sumCDcycles3 (\upi^C \cr \upi^D)_{\mbox{}_{\mbox{\scriptsize 3}}}/2 + V(-\shortcr-/-\shortcr-) = E  \mbox{ } . 
\label{E-Aff}
\eeq

\subsection{Some notions of 2-$d$ relational configuration also admit a $\mathbb{C}$ formulation}\label{Complex}

As well as the two components of $Tr(2)$ being an obligatory pairing in this setting (keep both or none), 
$Rot(2)$ and $Dil$ are also an obligatory pairing: as the modulus and phase parts of a single complex number.

The simplest notions of relational configuration that admit a $\mathbb{C}$ formulation have configuration spaces forming the diamond array 
$\mathbb{C}^N$, $\mathbb{C}^n$, $\mathbb{CP}^n$, $\mathbb{CP}^{n - 1}$ corresponding to quotienting out by none, one or both of $Tr(2)$ and $Rot(2) \times Dil$

In one sense, complex representability extends to the 2-$d$ affine case: its Procrustean stretch and shear can be co-represented in complex form
On the other hand, this is a $\gamma \bar{z}$ representation, i.e. antiholomorphic, 
and furthermore area is not particularly natural in the complex plane, since it is an Im part rather than a whole complex entity: $\mbox{Im}(z^A \bar{z}^B)$.

For sure, M\"{o}bius Configurational Relationalism continues to lie firmly within the complex domain. 

{            \begin{figure}[ht]
\centering
\includegraphics[width=0.4\textwidth]{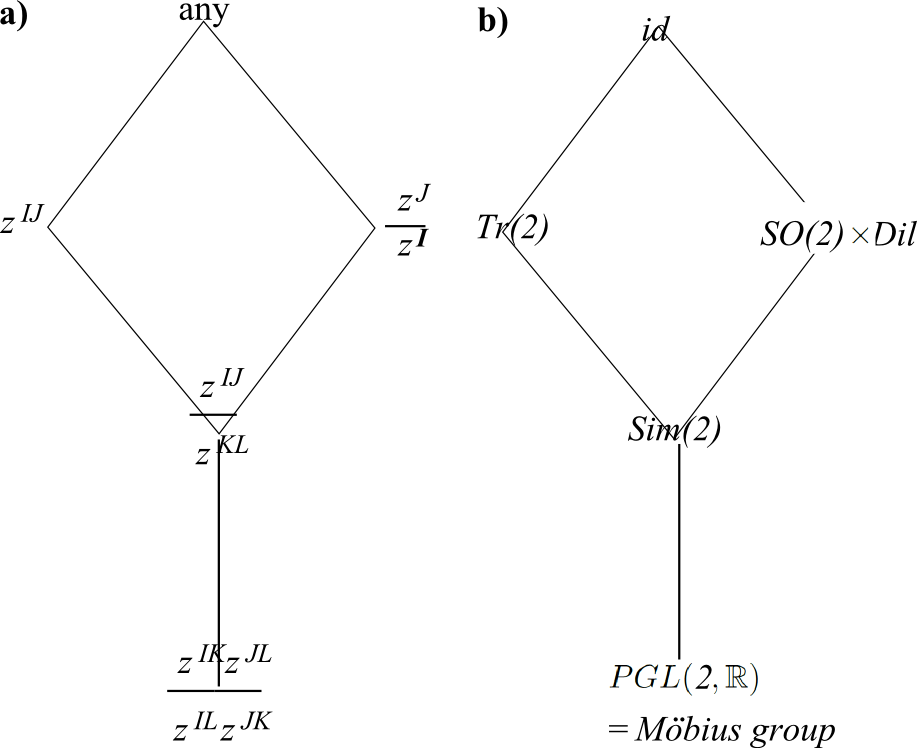}
\caption[Text der im Bilderverzeichnis auftaucht]{        \footnotesize{ Complex suite of a) invariants and b) the corresponding groups $\FrG$. 
The M\"{o}bius group has further subgroups amenable to complex formulation that are not considered here.} }
\label{Complex-Suite} \end{figure}          }

Complex examples of Best Matching include the following. 
The translation correction is $\d A = \d a_x + i\,\d a_y$ and the rotation-and-dilation correction is 
\beq 
\d B \, z^I               \mbox{ } \mbox{ for } \mbox{ }  \d B = \d c + i \, \d b \mbox{ } .
\eeq
Affine best matching -- new to this paper -- completes this with the Procrustean-and-shear correction          
\beq
\d C \, \bar{z}^I        \mbox{ } \mbox{ for } \mbox{ }  \d C = \d f + i \, \d e \mbox{ } . 
\eeq
On the other hand, M\"{o}bius best matching -- also new to this paper -- completes the above with a holomorphic quadratic correction: 
\beq
\d M \, \bar{z}^{I\,2}   \mbox{ } \mbox{ for } \mbox{ }  M \, \in \, \mathbb{C} \mbox{ } . 
\eeq
\ni The new geometrical entity of particular interest are the cross-ratio spaces $\FrC(N, 2)$,
In particular the minimal relationally nontrivial unit is $\FrC(4, 2)$.
Whereas each cross ratio is itself a complex number-valued quantity, it is not yet clear which geometrical structure is natural to cross-ratio space.

\subsection{Examples of complex Mechanics formulations}

The zero total momentum constraint, if present, is just ${\cal P}^z = \sumIN p_I$, for each $p_I$ of form $p_{xI} + i p_{yI}$.  
On the other hand, the complex {\it zero total dilational-and-angular momentum constraint} is (new to this paper)
\beq
{\cal Q}^z := \sumIN \bar{z}^Ip_I = 0 \mbox{ } .
\eeq
A complex action for $Sim(2)$ is then (using complex vector norm, and new to this paper,  
\beq
S = \sqrt{2}\int \mbox{\LARGE ||}  \frac{\d_{A, B}\bz}{\bz}  \mbox{\LARGE ||}  \sqrt{E - V} \mbox{ } , 
\eeq
for $V = V(-/-)$.  
The corresponding form of the quadratic primary constraint is
\beq
{\cal E} := ||\bz||^2||\bp||^2/2 + V(-/-) = E \mbox{ } . 
\eeq
\ni Finally, with $Conf(d)$ of course becoming infinite-dimensional in 2-$d$, 
I also briefly consider a finite Lie subgroup in that case, i.e. the M\"{o}bius group as a finite substitute. 
[An infinite group of physically irrelevant transformations would trivialize any finite particle configuration.]  
The {\it zero total M\"{o}bius momentum constraint} -- new to this paper -- is 
\beq
{\cal M}^z := \sumIN \bar{z}^{I\,2} p_I = 0 \mbox{ } .  
\eeq
Moreover, provision of an indirectly formulated action for this is blocked.
I.e. in this case, for now we reach a new impasse rather than a new theory of RPM. 
This occurs due to the following reason.  
Whereas one can readily enter changes into a ratio so as to produce a function that is homogeneous linear in change, 
in a cross ratio everything which enters in the numerator also features in the denominator. 
This forces the change to feature homogeneously with degree zero.

\section{Some research frontiers in Shape Mechanics}\label{Mech-Front}

\ni \underline{Frontier 1}. An important next step is to work out reduced versions of the RPM actions. 
For metric shape and scale-and-shape theories, this was covered in \cite{FORD, Cones, FileR}.
The GR counterpart of this procedure is also well known, in general leading to the impasse known as the Thin Sandwich Problem \cite{San-1, San-2}.  
Moreover, reduction is a means of getting at least some candidates for the (generalized) shape space geometries 
upon which to base the both the corresponding Shape Statistics and the geometrical reduced quantization scheme.

\ni \underline{Frontier 2}. Distinct direct considerations can on some occasions permit finding the relational configuration space and then building a Mechanics thereupon. 
E.g. this occurs for 1- and 2-$d$ metric shape RPM \cite{FORD, FileR}; moreover, in this case direct consideration and indirect consideration followed by reduction coincide.
To what extent does this coincidence extend among the new theories laid out in this paper?

\mbox{ }

\ni For instance, the direct implementation of M\"{o}bius RPM remains open in this case, 
due to the inherent problem with making a homogeneous linear function out of cross-ratios.
I.e. that converting any number of $z_i$ in $[z_1, z_2 ; z_3, z_4]$ into $\d w_i$ quantities does {\sl not succeed} in giving a quantity homogeneous linear in the $\d$'s.  

\mbox{ }

\ni \underline{Frontier 3}. 
Some projective geometries have point-to-line duality.  
An interesting question for the foundations of Mechanics then is if one sets up a `point particle Mechanics' here, can it be reinterpreted as a dual `line Mechanics' thereupon? 
How does this impact upon our preconceptions of classical dynamics?

\section{Shape Statistics} 

\subsection{Metric Shape Statistics}\label{MSS}

{\it Clumping Statistics} investigates hypotheses concerning ratios of relative separations (detailed information which can be attributed locally and to subsystems).   
These already exist in 1-$d$ and in settings simpler than metric shape spaces, so this topic is well-known.
Astrophysical situations modelled by this include tight binary stars, globular clusters, galaxies and voids: {\sl absense} of clumping.   
E.g. Roach \cite{Roach} provided a discrete statistical study of clumping; this can in turn be interpreted in terms of coarse-grainings of RPM configurations.
Also note that Geometrical Probability on the shape space $\FrS(N, 1) = \mathbb{S}^{n - 1}$ provides an alternative method to this.

\mbox{ } 

\ni Next, consider completing the above at the metric shape space level to probing relative angle information, the existence of which requires $d \geq 2$. 
E.g. Kendall \cite{Kendall84, Kendall89} investigated the relative angle question of whether the locations of the standing stones of 
Land's End in Cornwall contained more alignments than could be put down to to random chance.\footnote{Kendall's work is a {\sl solution} 
to Broadbent's \cite{Broadbent} previous posing of the standing stones problem as one to be addressed by {\sl some kind of} Geometrical Probability.}
%
This involves the following procedures.

\mbox{ } 

\ni 1) Sample in threes.  

\ni 2) Consider whether there were a statistically significant number of almost collinear triangles quantified by a bluntness angle $< \epsilon$ some small value (Fig \ref{Probes}.a).

\ni 3) Use probability distributions based on the corresponding shape space geometry (i.e. on Kendall's spherical blackboard).  

\mbox{ }

\ni Another application of Relative Angle Statistics is disproving claims of quasar alignment.\footnote{Moreover for `observed sky' applications, 
the space of {\sl spherical} triangles \cite{Kendall89} is even better. 
See \cite{Kendall} for a review of the corresponding shape geometry and Shape Statistics, 
and \cite{ASphe} for the related topic of Shape Mechanics built from stripping down $\mathbb{S}^d$ rather than $\mathbb{R}^d$ absolute spaces.}
%
Shape Statistics has also been applied to biological modelling of spoecific 3-$d$ objects (e.g. skulls) 
viewed as ratios and relative angles based upon some approximating collection of `marker points' \cite{Small}. 

\mbox{ } 

\ni Significant results for different values of $\epsilon$ carry different implications \cite{Broadbent}. 
Were the standing stones laid out skillfully by the epoch's standards for e.g. astronomical or religious reasons ($\epsilon \leq 10$ minutes of arc), 
or were they just the markers of routes or plots of land ($\epsilon \leq 1$ degree)?

\mbox{ }

\ni I also point out here that metric Shape Statistics (whether or not scaled) is likely to be a useful tool for Robotics. 
For instance, this can be used to analyze the extent to which robots adopt approximately the same configuration in response to similar external conditions. 
Or as regards the extent to which there is success in robots mimicking (sequences of) animal configurations (fall like a cat, run like a horse...).
I.e. (toy model) robotic configuration spaces not only involve determining the path properties outlined in Sec \ref{Q-Geom}, 
but are also the basis for theories of firstly Geometrical Probability and secondly of Shape Statistics. 

\mbox{ } 

\ni \underline{Frontier 4}. As regards specific Mechanics questions to be settled by Shape Statistics, 
consider whether a given Celestial Mechanics configurations exhibit a shape statistically significant number of eclipses: 
clumping-based questions include whether it exhibits a shape statistically significant number of tight binaries. 
Whether it contains a shape statistically significant globular cluster. 
Finally, whether two images of globular clusters have sufficiently similar clustering detail to be of the same system or to be members of similarly formed populations.
%
{            \begin{figure}[ht]
\centering
\includegraphics[width=1.0\textwidth]{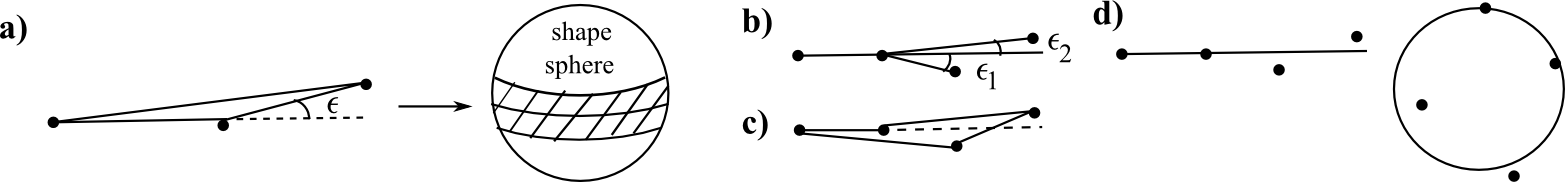}
\caption[Text der im Bilderverzeichnis auftaucht]{        \footnotesize{a) $\epsilon$-bluntness in probing metric shapes. 
b) and c) are assessments of almost-colinearity in fours in the conformal and affine settings. 
In the M\"{o}bius setting, cross ratios are to be used in this assessment and the method has to consider almost-cocircularity in fours alongside almost collinearity in fours.} }
\label{Probes} \end{figure}          }

\mbox{ }  

\ni N.B. that the metric case of Shape Statistics is a well-established approach \cite{Kendall84, Kendall89, Small, Kendall}.  
Whereas the current paper concentrates on laying out many new theories of background-independent Mechanics, 
these are equally tied to being geometrical theories of shape with statistical applications.
If the reader wishes to see what these look like for metric shapes -- i.e. similarity-redundant geometry -- they should cast a look through \cite{Kendall}.  
In the current article, I point out there is an analogue of this where it is each of conformal, affine, equivoluminal groups that are redundant, 
or where these, the Euclidean and the similarity group are supersymmetrized.
It is on this basis that I point to a number of frontiers.
These are likely to be of interest to applied topologists, to geometers, to people working in Probability and Statistics, as well as to people working in Theoretical Mechanics 
and on the Foundations of Physics, in particular on models of Background Independence for use in Quantum Gravity.

\subsection{More general Shape Statistics}

I next entertain the idea of the above Shape Statistics being but the first of a larger family, each corresponding to a distinct notion of shape.
A more generalized methodology is as follows.

\mbox{ } 

\ni 1) Find the probe unit, as tabulated in the last three columns of Fig \ref{Tab-1}.

\ni 2) Find the geometry of the space of the probe unit (the next paper in this series); 
by the discussion in Sec \ref{Comparer} this extends to comparison of two 'best fit' configurations.

\ni 3) Build geometrical probability theory thereupon.
Note that this being geometrical is not always necessary since some shape spaces turn out to be flat (see \cite{AConfig} for examples).   

\ni 4) Build Shape Statistics -- using a restricted region of 2) -- corresponding to the probe unit taking some particular distinctive form.  

\mbox{ }

\ni For each Shape Statistics, one application is to the classical Records Theory of the corresponding Shape Mechanics.

\subsection{Affine and equi-top-form Shape Statistics}

\underline{Frontier 5}. One likely application of the 2-$d$ case -- Area and Area Ratio Statistics -- is image recognition.
In looking at images of point configurations, or approximating images using sampling points, the flat space affine Shape Statistics of images only makes sense if probing at least in fours. 
The minimal equivalent of the spherical blackboard is the affine shape space $\FrA(4, 2)$, of dimension 2. 

\mbox{ }

\ni Our illustrative question then is whether Thompson's fish are affine shape statistically significantly coincident under affine transformations, 
or whether they `just look that way' much like the standing stones of Land's End `look to the eye' to have a lot of collinearities in threes.
For now I present a 2-$d$ analogue of the fish problem (as per Fig \ref{Matching-2}.b), which points to use of equiareal Shape Statistics in this case.

\subsection{Conformal Shape Statistics, alias Local Angle Statistics}

This concerns quantitative analysis of propositions concerning local angles exhibited by a point distribution. 
In particular in 3-$d$, the minimum sampling unit here is probing in sixes, corresponding to a 3-$d$ angle space $\FA(4, 3)$.

\subsection{M\"{o}bius Shape Statistics as a Cross-Ratio Statistics, and other such}

This is but one Cross-Ratio Statistics, since cross ratios are invariants in a wider range of projective geometries. 
In the present case in the complex plane, the minimum probing size is 4 points, corresponding to the 2-$d$ cross-ratio space $\FrC(4, 2)$.

\mbox{ } 

\ni As per Fig \ref{Probes}.b-c), this case has a very natural analogue of the Euclidean case's `collinearity in threes' hypothesis.
Firstly note that in this geometrical setting lines and circles can be mapped to each other, so collinear is to be upgraded to that, and {\sl four points} are needed. 
Secondly, with these changes made, the question becomes how many almost-real cross ratios there are.  
This makes use of the well-known fact that real cross ratios correspond to a line-or-circle passing through the four points, with `almost real' in the role of bluntness parameter.  

\mbox{ } 

\ni \underline{Frontier 6}. The above may be used to test whether a given picture was drawn according to the rules of perspective. 
Many pictures might na\"{i}vely {\sl look} like perspective drawings, but the standing stones problem has illustrated the gap between that and statistical significance. 
The above method can firstly be used to settling this issue, and secondly -- by using diverse tolerance parameters -- 
can be used to assess how accurately a confirmed perspective drawing was crafted. 
This could in turn be used as a means of dating the drawing, identifying the equipment used, 
or helping to identify the artist by comparison with the value of `almost real' in the paintings reliably attributed to them. 
Conversely, this method may have the capacity to spot wrongly attributed pictures and forgeries. 
Note that this can be considered to be another type of image analysis.

\subsection{Using multiple Shape Statistics, or one as general as needed}

\underline{Frontier 7}. Investigate whether some instances of evolving objects 
-- e.g the shape of a particular animal's skull as that animal ages, of skulls over the course of the evolution of species, or of galaxies --
preserve relative angle information to a greater extent than ratio information. 
Investigating whether this is significant to the extent of each such sequence involving conformal maps would require setting up a conformal Shape Statistics.  
Other hypotheses would include that assessing angle information within metric Shape Statistics suffices, or that ageing and evolution are better modelled affinely. 
For instance, what type of geometrical transformation is general enough to model each of the three examples given above? 

\mbox{ }

\ni In conclusion, having a Shape Statistics for {\sl each type} of geometry multiplies opportunities of spotting significant patterns in nature.

\section{Supersymmetric RPM}\label{Susy-RPM}

GR has a clear-cut Temporal and Configurational Relationalism split as laid out in Appendix C. 
On the other hand, Appendix D explains how Supergravity does not.  
This difference stems from Supergravity's constraint algebraic structure being more complicated than GR's. 
The above distinction between GR and Supergravity then has further knock-on effects as regards other aspects of Background Independence, 
such as 1) as regards which notions of observables remain meaningful.
2) Supergravity does not have a direct analogue of Superspace.\foo{For sure, this is meant here in Wheeler's sense, 
rather than in the entirely technically different sense used in the Supersymmetry literature.}
%
Moreover, Supersymmetry is itself a source of constraints, 
and the example of Supergravity shows that capable of overriding the importance of other sources of constraints such as Temporal Relationalism and Configurational Relationalism.
Due to these observations, whether it is possible for Supersymmetry to be compatible with Relationalism and with Background Independence more generally is an interesting question.  
Below, we settle this matter in the affirmative by constructing supersymmetric RPMs for the first time.  

\mbox{ } 

\ni Absolute space is here a $\mathbb{Z}_2$-graded version of $\mathbb{R}^d$: $\{\FA = \mathbb{R}^{(d|2p)}\}^N$ for $\mN = p$  Supersymmetry, 
meaning with $p$ supercharges each accompanied by conjugates.
The $\FA \rightarrow \times_{i = 1}^N \FA$ construct then continues to apply: $\FrQ(d|2p) = \mathbb{R}^{(Nd|2p)}$.
Then see Fig \ref{Susy-Groups} for eleven supersymmetric $\FrG$ which are subgroups of one or both of the superconformal and superaffine `super-apex groups'.
Each provides a corresponding notion of Relationalism and a reduced configuration space geometry.  
It is possible also at least in principle to consider a supersymmetric $\FrQ$ subject to a `merely bosonic' $\FrG$ (such as those tabulated in Fig \ref{More-G}).  

\mbox{ }

\ni Finite models including fermions attain Temporal Relationalism \cite{FileR} through being homogeneous linear geometries   
\beq
\d s = \sqrt{\mm_{\sfA\sfB}   \d{Q}^{\sfA} \d{Q}^{\sfB}}\sqrt{W}    + \ml_{\sfC} \d{Q}^{\sfC} \mbox{ } .  
\label{Schema}
\eeq
Note that the action (\ref{Schema}) is no longer of Jacobi type but rather of Randers type \cite{Randers} (a subcase of Finsler geometry if it is additionally nondegenerate, 
and of Jacobi--Synge type action).  
Here $\bm$ is a quadratic `bosonic' contribution to the overall notion of metric involved, and $\bl$ is a linear `fermionic' contribution.  
This is in the further context of the species indexed by $\fA, \fB$ on the one hand and by $\fC$ on the other are not to be overlapping in this setting 
(i.e. a partition into distinct bosonic and fermionic species respectively).  
(\ref{Schema}) is also a model arena for the relational form of Einstein--Dirac Theory \cite{Van}.  
The above presentation is prior to applying Best Matching with respect to the $\FrG$ in question.

\mbox{ } 

\ni Let us next apply that specifically in the case of $\mN = 1$, $d = 1$ and $\FrG = superTr(1)$. 
Here the particles indexed by $I$ each have coordinates $q^I, \theta^I, \bar{\theta}^I$, 
due to the usual 1-$d$ $x$ being accompanied by two Grassmann coordinates $\theta$ and $\bar{\theta}$.\footnote{In 1-$d$, the version with only one Grassmann coordinate is also possible.
The non-relational version of Supersymmetric Mechanics with two Grassmann coordinates was first studied by Nicolai \cite{N76} and Witten \cite{W81}.}
%
Then 
\beq
S_{\sss\su\sss\sy}[q^I, \theta^I, \bar{\theta}^I, a, \alpha] = \sqrt{2}\int\left\{ ||\d_{a,\alpha}\bq||\sqrt{W} + i \sumIN\{\bar{\theta}^I\d_{a,\alpha}\theta^I - \overline{\d_{a,\alpha}\theta}^I\theta^I \}\right\}
\eeq
for fermionic best-matched derivatives
\beq
\d_{a,\alpha}\theta^I := \d\theta^I - \d a + i \d \alpha \mbox{ } ,  \mbox{ }  \mbox{ } \overline{\d_{a,\alpha}\theta}^I := \d\bar{\theta}^I - \d a - i\d\bar{\alpha} 
\eeq
and bosonic best-matched derivatives
\beq
\d_{a,\alpha}q^I := \d q^I - \d a - \bar{\theta}^I\d\alpha - \d\bar{\alpha} \, \theta^I \mbox{ } .  
\eeq
The $\d\alpha$ and $\d\bar{\alpha}$ corrections to the fermionic species are `Grassmann translations'. 
Furthermore, upon imposing Supersymmetry these also feature as corrections to the bosonic changes in the Grassmann-linear manner indicated. 
%

Then variation with respect to $a$ gives a new form of 1-$d$ zero total momentum of the universe constraint 
\beq
{\cal S}_{\sss\su\sss\sy} := \sumIN \{p_I + p_{\theta I} - p_{\bar{\theta} I} \} = 0 \mbox{ } .  
\eeq 
The new form just reflects that fermions also carry momentum.
On the other hand, variation with respect to $\alpha$ and $\bar{\alpha}$ give the {\it zero total supersymmetric exchange momentum of the universe} constraints
\beq
{\cal S}           := - \sumIN \big\{ p_{\theta I}       + i \bar{\theta^I}p_I \big\} = 0 \mbox{ }, \mbox{ } \mbox{ } 
{\cal S}^{\dagger} :=   \sumIN \big\{ p_{\bar{\theta} I} + i      \theta^I p_I \big\} = 0 \mbox{ } . 
\eeq
Note that these gain one piece from the fermionic sector and one piece from the bosonic sector.
These constraints are accompanied by the standard quasi-bosonic ${\cal E}$, except that now $V$ contains fermionic species also: 
\beq
{\cal E} := ||\mbox{\boldmath{$p$}}||^2/2 + V(q^I, \theta^I, \bar{\theta}^I) = E \mbox{ } .  
\eeq
Taking for now the stance of not knowing the supersymmetric analogues of shape, the incipient form of $V$ is
\beq
V(q^I, \theta^I, \bar{\theta}^I) = V_B(q^K) + \sumIN \left\{ \theta^I u_I(q^K) - \bar{\theta}^I v_I(q^K) + \sumJN\theta^I\bar{\theta}^J w_{IJ}(q_K) \right\} \mbox{ } ,     
\eeq
by virtue of the automatic truncation in Grassmann polynomials afforded by the underlying anticommutativity. 
Then demanding algebraic closure gives the conditions on $V$ for $V$ to be a function of the $superTr(1)$ notion of shape as
\beq
\sumIN \bar{\theta}^I \left\{  \frac{\pa V_B(q^{K})}{\pa q^I} + \sumJN     \theta^J  \frac{\pa u_J(q^{K})}      {\pa q^I}  \right\} = 0 \mbox{ }, \mbox{ } \mbox{ }
\sumIN      \theta^I  \left\{ -\frac{\pa V_B(q^{K})}{\pa q^I} + \sumJN\bar{\theta}^J \frac{\pa \bar{v}_J(q^{K})}{\pa q^I}  \right\} = 0 \mbox{ } .
\label{Susy-Shape}
\eeq
\ni \underline{Frontier 8}. Gain an understanding of the notion of `supershapes'.

\mbox{ } 

\ni Furthermore, the above implementation of best matched Configurational Relationalism readily extends to all dimensions and, concurrently, 
to apply to all the other supersymmetric $\FrG$ listed whose Supersymmetry is tied to the momentum generators $P_i$. 

\mbox{ } 

\ni \underline{Frontier 9}. Work in ecess to that presented here is required for those theories in which Supersymmetry is tied to special conformal transformations $K_i$ instead, 
with the principal remaining question of interest being the production of an explicit superconformal RPM.  

\mbox{ } 

\ni The above implementation proceeds firstly by correcting terms with spatial vector indices $\d q^{Ii}$ by subtracting products of\footnote{Dotted and undotted Greek indices here are 
a standard spinorial index notation.} 
$\bar{\theta}^{I\dot{\alpha}}$ and $\d\alpha^{\alpha}$, and of $\theta^{I\alpha}$ and $\d\bar{\alpha}^{\dot{\alpha}}$, 
necessitating `interconversion arrays' ${A^i}_{\alpha\dot{\alpha}}$.
Then familiarity with standard spinorial formulations points to these arrays being e.g. the vector of Pauli matrices in dimension 3, and corresponding generalizations 
in further dimensions based on that dimension's corresponding Dirac and Clifford mathematics. 
This case also requires use of the well-known distinction between $^\dagger$ and $\bar{\mbox{ }}$.  
Additionally, it is clear from super-brackets relations at the level of the algebra that fermionic species are {\sl rotational} spinors and also carriers of nontrivial homothetic weight.
These give corresponding matching corrections to the fermionic changes. 
Upon variation with respect to the translational and rotational auxiliaries, the preceding two best matching terms contribute, respectively, fermionic angular momentum 
to the zero total angular momentum constraint and fermionic dilational momentum to the zero total dilational momentum constraint.
In the $SL(d, \mathbb{R})$ case, fermionic species are indeed  $SL(d, \mathbb{R})$ spinors, giving corresponding matching conditions, 
and contributing fermionic angular momentum, shear and Procrustean stretch to the corresponding zero total angular momentum, shear and Procrustean stretch constraints.

\mbox{ } 

\ni The given $superTr(1)$ theory suffices to establish that Supersymmetry 
is another setting in which a simple removal of the centre of mass by passing to relative coordinates is not possible.
Nor does preemptively taking out the centre of mass allow for one's philosophical worldview and subsequent physical paradigm to avoid the possibility of Supersymmetry. 
This is since {\sl relative} translations still exist within that setting, 
and these are sufficiently similar mathematically to absolute translations to enable Supersymmetry in the usual manner.

\mbox{ } 

\ni \underline{Frontier 10}. Understand the extent to which superRPMs can be cast in reduced form.  

\ni \underline{Frontier 11}. Whether directly or by the preceding reduction, what is the topology and geometry of each $\FrG$'s super-shape space? 

\mbox{ } 

\ni [The theory of supershapes is not well-known enough for questions about `supersymmetric Shape Statistics' to be posed for now.] 

\mbox{ } 

\ni \underline{Frontier 12}. The range of supersymmetric RPMs sketched out in the current paper is of anticipated future value in 
assessing the extent to which Relationalism and other aspects of Background Independence can be reconciled with Supersymmetry. 
This is in anticipation of more detailed investigation of how the former are substantially altered in passing form GR to Supergravity \cite{ABeables}.  

\mbox{ } 

\ni Finally N.B. that in the above theories 
\beq
\mbox{\bf \{}{\cal S}\mbox{\bf ,} \, \bar{{\cal S}}\mbox{\bf \}} \sim {\cal P} \mbox{ } \mbox{ and not } \mbox{ } {\cal E} \mbox{ } , 
\eeq
signifying that these supersymmetric RPMs are {\sl not} models of Supergravity's principal alteration (Appendix D)
as compared to GR (Appendix C) as regards aspects of Background Independence and subsequent Problem of Time facets. 
This is relevant to the discussion of differences between GR and Supergravity in Appendices C and D.
In particular, at least the super-RPM arena in which the Supersymmetry is tied to $P_i$ is one in which Configurational Relationalism can indeed include Supersymmetry,  
and do so without interfering with the `usual' separate provision of Temporal Relationalism.

\section{Quantum RPMs}\label{Quantum}

All of this paper's relational theories of Mechanics make for interesting quantum schemes.  
In each case, if a model is relational to this extent, how is the corresponding QM affected? 
It is not that different \cite{FileR} for metric RPM with and without scale!  
This is due to relative angular momentum (and relative dilational momentum \cite{AF}, and mixtures \cite{+Tri, QuadI}) having the same mathematics as angular momentum. 
Moreover, the quantum metric shape quadrilateral \cite{QuadIII} did produce a more unusual and distinctive combination of features of the Periodic Table and of Gell-Mann's eightfold way.
Whereas the conformal case is well known for one absolutist particle and in QFT setting, it is not known as an $N$-body relational problem as posed here. 

\mbox{ } 

\ni I make use of Dirac quantization, much as Smolin \cite{Smolin} did for the original RPM \cite{BB82}.  
Within Dirac quantization, one works with the standard position coordinates or Jacobi coordinates based kinematical quantization \cite{I84} 
(first done by Rovelli \cite{Rovelli} for the original RPM).
[In contrast, kinematical quantization becomes a nontrivial geometrical issue in reduced quantization, 
though understanding the geometry and topology of the reduced configuration space in question leads to this being resolved also.]
I also make use of the Laplacian operator ordering, which in the current flat redundant configuration space case is equivalent also to the conformal operator ordering and the 
$\xi$ operator orderings more generally.
[These differ from the Laplacian by $- \xi R$ for $R$ the Ricci scalar of the configuration space.
A particular configuration space dimension dependent value of $\xi$ renders the overall operator conformally invariant, hence constituting the conformal operator ordering.]
These operator orderings were originally proposed by DeWitt \cite{DeWitt57} for use in what became GR Quantum Cosmology through the pioneering works of Misner \cite{Magic} 
(see also \cite{+Op} for uses of such operator orderings).

\mbox{ } 

\ni Example 1) The metric scale-and-shape RPM has   
\beq
\underline{\widehat{{\cal L}}}\Psi = \mbox{$\frac{\hbar}{i}$} \sumAn \urho^A \cr \mbox{$\frac{\pa\Psi}{\pa\usrho^A}$} = 0 \mbox{ } , 
\label{L-hat}
\eeq
meaning that $\Psi = \Psi(-\cdot-)$.  
The `main wave equation' is  
\beq
\widehat{{\cal E}}\Psi = -\mbox{$\frac{\hbar^2}{2}$} \triangle_{\mathbb{R}^{nd}} \Psi + V(-\cdot-)\Psi = E \Psi \mbox{ } , 
\label{E-hat}
\eeq
(in this case, conformal operator ordering collapses to Laplacian ordering since $\mathbb{R}^{nd}$ is flat).  
This example is but a slight upgrade of Smolin's \cite{Smolin} (which involved particle coordinates rather than Jacobi coordinates).

\mbox{ } 

\ni Example 2) Metric shape RPM has (\ref{L-hat}) and \cite{FileR} 
\beq
\widehat{{\cal D}}\Psi = \mbox{$\frac{\hbar}{i}$}\sumAn \urho^A \cdot \mbox{$\frac{\pa\Psi}{\pa\usrho^A}$} = 0 \mbox{ } ; 
\label{D-hat}
\eeq
together, these mean that $\Psi = \Psi( -\cdot-/-\cdot-)$.  
The `main' wave equation can then be expressed as
\beq
\widehat{{\cal E}}\Psi = -\mbox{$\frac{\hbar^2}{2}$} \triangle_{\mathbb{R}^{nd}} \Psi + V(-\cdot-/-\cdot-)\Psi = E\Psi \mbox{ } 
\label{E-hat-2}
\eeq
by use of conformal flatness of the configuration space metric.  

\mbox{ } 

\ni Example 3) New to the current paper, 3-$d$ Conformal shape RPM has 
\beq
\widehat{{\cal K}}_i\Psi = \mbox{$\frac{\hbar}{i}$}\sumIN \{ q^{I\,2} {\delta_i}^j - 2q_iq^j \}\mbox{$\frac{\pa\Psi}{\pa q^{iI}}$} = 0 \mbox{ } , 
\label{K-hat}
\eeq
which in conjunction with
\beq
\widehat{\underline{{\cal P}}}\Psi = \mbox{$\frac{\hbar}{i}$} \sumIN \mbox{$\frac{\pa\Psi}{\pa \underline{q}^{I}}$}
\label{P-hat}
\eeq
and the $\bq$-versions of (\ref{L-hat}) and (\ref{D-hat}) signify that $\Psi = \Psi(\angle)$.
The main wave equation is then 
\beq
\widehat{{\cal E}}\Psi = - \mbox{$\frac{\hbar^2}{2}$} I \, \triangle_{\mathbb{R}^{Nd}} \Psi + V(\angle)\Psi = E\Psi \mbox{ } .  
\label{E-hat-3}
\eeq
\ni Example 4) Also new to the current paper, 2-$d$ Area RPM has
\beq
\widehat{\underline{{\cal S}}}\Psi = \mbox{$\frac{\hbar}{i}$} \sumAn \urho^A \underline{\underline{\underline{S}}} \mbox{$\frac{\pa\Psi}{\pa\usrho^A}$} = 0 \mbox{ } .  
\label{S-hat}
\eeq
This means that $\Psi = \Psi(-\shortcr-)$.
Noting that this and the next two examples lie outside the levels of structure upon which Laplacians and conformal Laplacians are defined, 
so it is even less clear for now in these cases how to operator order, the `main wave equation' for 4-particle area RPM is 
\beq
\widehat{{\cal E}}\Psi = -\mbox{$\frac{\hbar^2}{2}$} \sumABcycles3 \big( \mbox{$\frac{\pa}{\pa\urho^{A}}$} \cr \mbox{$\frac{\pa\Psi}{\pa\urho^{B}}$})_{\mbox{}_{\mbox{\scriptsize 3}}} + V(-\shortcr-) \Psi = E \Psi \mbox{ } . 
\label{E-hat-4}
\eeq
\ni Example 5) New again to the current paper, 2-$d$ affine shape RPM has linear constraints (\ref{S-hat}) and (\ref{D-hat}), or, formulating them together, 
\beq
\widehat{\underline{{\cal G}}}\Psi = \mbox{$\frac{\hbar}{i}$}   \sumAn \urho^A \underline{\underline{\underline{G}}} \, \mbox{$\frac{\pa\Psi}{\pa\usrho^A}$} = 0 \mbox{ } .  
\label{G-hat}
\eeq
This means that $\Psi = \Psi(-\shortcr-/-\shortcr-)$.
The main wave equation for 4-particle affine shape RPM is 
\beq
\widehat{{\cal E}}\Psi = -\mbox{$\frac{\hbar^2}{2}$} \sumCDcycles3(\urho^{C} \cr \urho^{D})_{\mbox{}_{\mbox{\scriptsize 3}}} 
                                                 \sumABcycles3(\mbox{$\frac{\pa}{\pa\urho^{A}}$} \cr \mbox{$\frac{\pa\Psi}{\pa\urho^{B}}$})_{\mbox{}_{\mbox{\scriptsize 3}}}\Psi 
					 + V(-\shortcr-/-\shortcr-)\Psi = E\Psi \mbox{ } . 
\label{E-hat-5}
\eeq

\ni Example 6) As a final new Dirac quantization in this paper, in the case of $superTr(1)$ RPM, 
\beq
\widehat{{\cal S}}           = -\mbox{$\frac{\hbar}{i}$}\sumIN \big\{ \mbox{$\frac{\pa}{\pa\theta^I}$} + i\bar{\theta}^I\mbox{$\frac{\pa}{\pa q^I}$} \big\} \Psi = 0 \mbox{ } ,
\label{Q-hat}
\eeq
\beq
\widehat{{\cal S}}^{\dagger} =  \mbox{$\frac{\hbar}{i}$}\sumIN \big\{ \mbox{$\frac{\pa}{\pa\bar{\theta}^I}$} + i\theta^I\mbox{$\frac{\pa}{\pa q^I}$} \big\} \Psi = 0  \mbox{ } ,
\label{Q-dagger-hat}
\eeq
\beq
\widehat{{\cal P}}_{\sss\su\sss\sy}\Psi = \mbox{$\frac{\hbar}{i}$}\sumIN \big\{ \mbox{$\frac{\pa}{\pa q^I}$} + \mbox{$\frac{\pa}{\pa\theta^I} 
                                                                                                         - \mbox{$\frac{\pa}{\pa\bar{\theta}^I}$}$} \big\} \Psi = 0 \mbox{ } ,
\label{superP-hat}
\eeq
\be
\widehat{{\cal E}} \Psi-\mbox{$\frac{\hbar^2}{2}$}\triangle_{\mathbb{R^N}}\Psi + V(q^I, \theta^I, \bar{\theta}^I)\Psi = E\Psi \mbox{ } .  
\label{E-hat-7}
\ee
with $V$ within the form allowed by (\ref{Susy-Shape}).  

\mbox{ } 

\ni \underline{Frontier 13}. Study this paper's new theories' Dirac quantization schemes.

\ni \underline{Frontier 14}. Obtain the reduced quantization schemes, from first obtaining the underlying corresponding geometry and then proceeding along the lines of 
i)  Isham's geometrical kinematical quantization and 
ii) formulation and solution of the wave equations.

\mbox{ }

\ni See Appendices C and D for this paper's final Frontiers.

\mbox{ } 

\ni{\bf Acknowledgements} To those close to me gave me the spirit to do this.  
And with thanks to those who hosted me and paid for the visits: Jeremy Butterfield, John Barrow and the Foundational Questions Institute.
Thanks also to Chris Isham, Julian Barbour and Niall \'{o} Murchadha for a number of useful discussions over the years, and to the Anonymous Referee for useful comments.  

\begin{appendices}

\section{Flat $\mathbb{R}^d$ geometries}\label{Flat-Geom}

\subsection{Real geometries}\label{Real-Geom}

I develop this here from a simple Kleinian position -- based on invariants corresponding to transformation groups -- 
by considering $\FrG \leq Aut(\langle \mathbb{R}^d, \sigma\rangle)$ for various layers of mathematical structures $\sigma$.\footnote{See \cite{Coxeter} for an in-depth account 
of the foundations of geometry, albeit not based on group theory.
See also \cite{Stillwell} for comparison between four approaches to the foundations of geometry, including the group-theoretic approach.
Finally, I use $\langle \mbox{ } , \mbox{ } \rangle$ to demarcate a space (prior to the comma) that is equipped with further structures (after the comma).}
%
$\sigma$ could be $\cdot$     (scalar products, i.e. the Euclidean metric $\delta_{ij}$), 
         but also $/$                         denoting ratios, 
		          $-$                         denoting differences, as feature e.g. in the Euclidean notion of distance, or
				  $\angle$                    denoting angles.
$\sigma$ could also be $\mbox{\Large$\wedge$}$: the top form wedge product supported in dimension $d$, 
e.g. area built out of cross products $\shortcr$ in 2-$d$ or volumes built out of scalar triple products $[\cr\cdot \mbox{ } ]$ in 3-$d$.  
Some geometries additionally allow for a number of combinations of these structures; see in particular column 1 of Fig \ref{Tab-1}.

To be clear about the above shorthands' definitions, let $\underline{u}$, $\underline{v}$, $\underline{w}$, $\underline{y}$	$\in \mathbb{R}^d$.
Then the scalar product is a 2-slot operation $\underline{u} \cdot \underline{v}$. 
The Euclidean norm alias magnitude is then a special case of the square root of this: $||\underline{v}|| := \sqrt{\underline{v} \cdot \underline{v}}$.
Also
\beq
\mbox{( {\it Euclidean distance} between $\underline{u}$ and $\underline{w}$ )} \mbox{ } := \mbox{ } ||\underline{u} - \underline{w}|| \mbox{ } ,
\eeq
i.e. the Euclidean norm of the difference between the two vectors $\underline{u} - \underline{w}$.
Ratio is then a 2-slot operation acting on scalars, e.g. a ratio of two components of vectors  
\beq
\mbox{( ratio of magnitudes of $\underline{u}$ and $\underline{w}$ )} \mbox{ } := \mbox{ }  \frac{||\underline{u}||}{||\underline{w}||} \mbox{ } , 
\label{Mag-Ratio}
\eeq
\beq
\mbox{( ratio of distances )}  \mbox{ } := \mbox{ } \frac{||\underline{u} - \underline{v}||}{||\underline{w} - \underline{y}||} \mbox{ } , 
\label{Dist-Ratio}
\eeq
\beq
\mbox{( ratio of scalar products )} \mbox{ } := \mbox{ } \frac{(\underline{u}\cdot\underline{v})}{(\underline{w}\cdot\underline{y})} \mbox{ } . 
\label{Dot-Ratio} 
\eeq

The {\it angle} between $\underline{u}$ and $\underline{w}$ is then the arccos of the particular combination 
\ni\beq
\mbox{( scalar product of unit vectors $\hat{\underline{u}}$ and $\hat{\underline{v}}$ ) } \mbox{ } = \mbox{ } (\hat{\underline{u}} \cdot \hat{\underline{v}})  
                                                                                         =  \frac{(\underline{u}\cdot\underline{v})}{||\underline{u}|| \,  ||\underline{v}||} \mbox{ } , 
																						 \label{Angle}
\eeq
which is a product of square roots of 2 subcases of (\ref{Dot-Ratio}).  
Finally, the $d$-volume top form is
\beq
\mbox{( areas of parallelograms formed by vectors $\underline{u}$, $\underline{v}$ ) } \mbox{ } = \mbox{ }
(\underline{u}\shortcr\underline{v})_{\mbox{}_{\mbox{\scriptsize 3}}} \mbox{ } \mbox{ in 2-$d$ } , 
\label{Area)}
\eeq
and 
\beq 
\mbox{( volumes of parallelepipeds formed by vectors $\underline{u}$, $\underline{v}$, $\underline{w}$ ) } \mbox{ } = \mbox{ }
[\underline{u}\shortcr\underline{v}\cdot\underline{w}] \mbox{ } \mbox{ in 3-$d$ } . 
\label{Vol}
\eeq
\ni Then possible $\FrG$ include the following; cases whose corresponding geometry is well-known are indicated.
See Figs \ref{Flat-Geom} and \ref{Ref-Inv} for the meanings of the types of transformations involved,\footnote{The {\it semidirect product group} $\nFrG = \FrN \rtimes \FrH$ is given by 
$(n_1, h_1)\circ(n_2, h_2) = (n_1 \varphi_{h_1}(n_2), h_1\circ h_2)$ for $\FrN \lhd \nFrG$: `$\FrN$ is a normal subgroup of $\nFrG$', $\FrH$ a subgroup of $\nFrG$ 
                                                                     and $\varphi:\FrH \rightarrow \mbox{Aut}(\FrN)$ a group homomorphism.
Compare the direct product's $(g_1, k_1)\circ(g_2, k_2) = (g_1 \circ       g_2,  k_1\circ k_2)$: this has no normal group specification, and trivial automorphism. } 
%
and columns 1 and 2 of Fig \ref{Tab-1} for a summary. 
$\FrG = id$: a trivial limiting case corresponding to no transformations being available.
$\FrG = Aut(\langle\mathbb{R}^d, -\rangle) = Tr(d)$:     {\it translations}                      $\underline{x} \rightarrow \underline{x} + \underline{a}$, 
                                                                                                 which form a $d$-dimensional Abelian group $\langle\mathbb{R}^{d}, +\rangle$.
$\FrG = Aut(\langle\mathbb{R}^d, /\rangle) = Dil$:      {\it dilations} alias {\it homotheties} $\underline{x} \rightarrow k\underline{x}$, 
                                                                                                 which form a 1-$d$ Abelian group $\langle\mathbb{R}^+, \cdot\rangle$.  
$\FrG = Aut(\langle\mathbb{R}^d, -/-\rangle) = Tr(d) \rtimes Dil$.
$\FrG = Aut(\langle\mathbb{R}^d, \cdot\rangle) = Rot(d)$: {\it rotations}                        $\underline{x} \rightarrow \underline{\underline{B}}\underline{x}$
forming the {\it special orthogonal group} $SO(d) := \{B \in GL(d, \mathbb{R}) \, | B^{\sT}B = \mathbb{I}, \mbox{det}B = 1\}$, which is of dimension $d\{d - 1\}/2$
$\FrG = Aut(\langle\mathbb{R}^d, -\cdot - \rangle) := Isom(\mathbb{R}^d) = Tr(d) \rtimes Rot(d) =: Eucl(d)$: the $d\{d + 1\}/2$-dimensional {\it Euclidean group} of isometries, 
                                                                                                   corresponding to {\it Euclidean geometry} itself.
$\FrG = Aut(\langle\mathbb{R}^d, \cdot/\cdot \rangle) = Rot(d) \times Dil$.
$\FrG = Aut(\langle\mathbb{R}^d, -\cdot-/-\cdot-\rangle) = Tr(d) \rtimes \{Rot(d) \times Dil\} =: Sim(d)$: the $d\{d + 1\}/2 + 1$ dimensional {\it similarity group} 
                                                                                                   corresponding to {\it similarity geometry}.
																								   																			  																				  
$\FrG = Aut(\langle\mathbb{R}^d, \mbox{\Large$\wedge$}\rangle) = SL(d, \mathbb{R})$:                the $d^2 - 1$ dimensional special linear group, 
                                                                                                    consisting of the $d\{d - 1\}/2$ rotations, $d\{d - 1\}/2$ 
																								    shears and $d - 1$ `Procrustean stretches'.  
$\FrG = Aut(\langle\mathbb{R}^d, \mbox{\Large$\wedge$}-\rangle) = Tr(d) \rtimes SL(d, \mathbb{R})$: the $d\{d + 1\} - 1$ dimensional `equi-top-form group' 
                                                                                                    corresponding to `equi-top-form geometry'
                                                                                                    (for $d = 2$, $\mbox{\Large$\wedge$}$ = $\cr$ 
																									and this is the quite well known {\it equiareal geometry}).  
$\FrG = Aut(\langle\mathbb{R}^d, \mbox{\Large$\wedge$}/\mbox{\Large$\wedge$}\rangle) 
                                                                     = GL(d, \mathbb{R})$:          the $d^2$-dimensional general linear group, 
                                                                                                    consisting of rotations, shears and Procrustean stretches now alongside dilations.  
$\FrG = Aut(\langle\mathbb{R}^d, (\mbox{\Large$\wedge$}-)/(\mbox{\Large$\wedge$}-)\rangle) 
                                                       = Tr(d) \rtimes SL(d, \mathbb{R}) =: Aff(d)$ the $d\{d + 1\}$-dimensional {\it affine group} of linear transformations, 
                                                                                                    corresponding to {\it affine geometry}.

So far, the above transformations can all be summarized within the form of the eq at the top of Fig  \ref{Flat-Geom-Fig}
The most general case included here is affine geometry, within which all the other $\FrG$ above are realized as subgroups.
%
{            \begin{figure}[ht]
\centering
\includegraphics[width=0.55\textwidth]{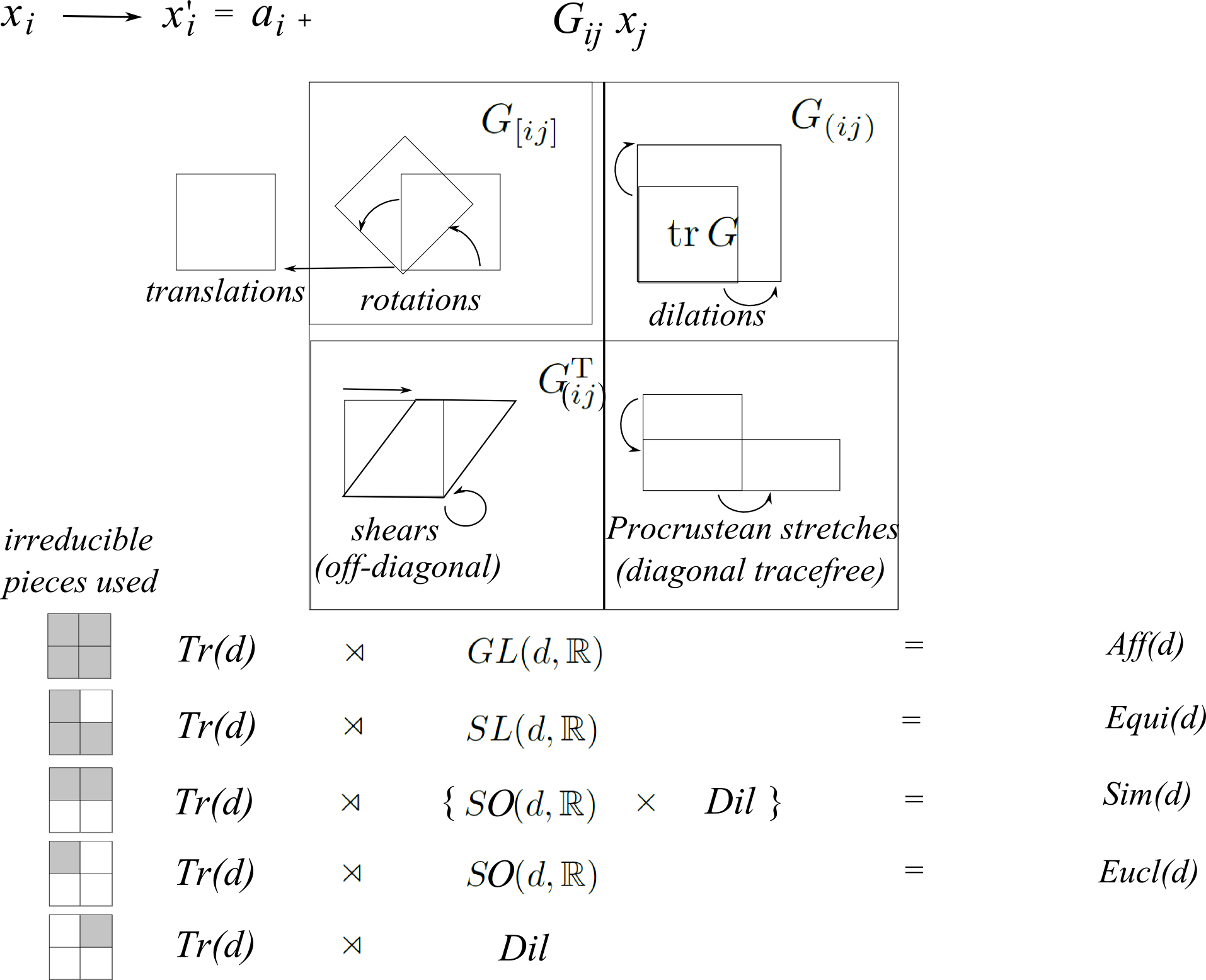}
\caption[Text der im Bilderverzeichnis auftaucht]{        \footnotesize{Elementary transformations. 
2-$d$ illustration of translation, rotation, dilation, shear, and Procrustean stretch 
(i.e. $d$-volume top form preserving stretches, in particular area-preserving in 2-$d$ and volume-preserving in 3-$d$). 
I also indicate the relation of the last four of these to the irreducible pieces of the general linear matrix $\underline{\underline{G}}$, 
and which geometrically illustrious groups these transformations form part of.
The T superscript denotes `tracefree part'.  
Note that Procrustean stretches do not respect ratios and shears do not respect angles.} }
\label{Flat-Geom-Fig} \end{figure}          }

\mbox{ } 

\ni {\it Reflections} could also be involved in each case. 
These are a third elementary type of isometry about an invariant mirror hyperplane (line in 2-$d$, plane in 3-$d$).
Unlike translations and rotations, they are a discrete operation.
For a mirror through the origin, characterized by a normal $\underline{n}$, the explicit form for the corresponding reflection is the linear transformation

\ni\beq
Ref: \underline{v} \rightarrow \underline{v} - 2 \, (\underline{v}\cdot\underline{n}) \, \underline{n} \mbox{ } . 
\eeq

Moreover, a further direction in $d$-dimensional geometry can be taken by introducing {\it inversions} in $\mathbb{S}^{d - 1}$, 

\ni\beq
Inv: \underline{v} \rightarrow \frac{\underline{v}}{||\underline{v}||^2} \mbox{ } .
\label{Inv}
\eeq
These {\sl also} preserve angles  -- but {\sl not} other ratios of scalar products  (Fig \ref{Ref-Inv}.b) --  
paving the way to the yet larger group of transformations that specifically preserves just angles.
%
{            \begin{figure}[ht]
\centering
\includegraphics[width=0.55\textwidth]{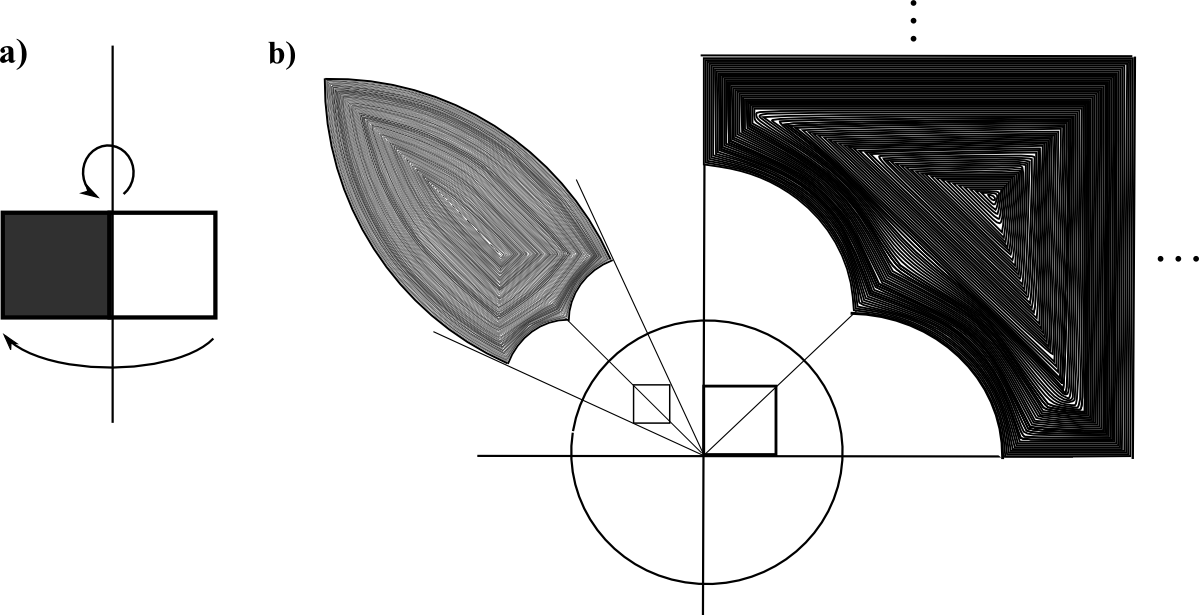}
\caption[Text der im Bilderverzeichnis auftaucht]{        \footnotesize{2-$d$ renditions of a) reflection, which in this case is about a mirror line. 
b) Inversion in the circle.  
This transformation requires a grid of squares to envisage -- rather than a single square -- since it has a local character which differs from square to square.  
N.B. also that this can map between circles and lines, with the sides of the squares depicted often mapping to circular arcs.} }
\label{Ref-Inv} \end{figure}          }

Another perspective on geometry involves weakening the five axioms of Euclidean geometry \cite{Coxeter, Stillwell}.
The best-known such weakening is {\it absolute geometry}, which involves {\sl dropping just} Euclid's parallel postulate. 
This leads firstly to hyperbolic geometry arising as an alternative to Euclid's, and then more generally to such as Riemannian differential geometry.  
In contrast with this, affine geometry is that this {\sl retains} Euclid's parallel postulate, and indeed places central importance upon developing its consequences (`parallelism').
This approach drops instead Euclid's right-angle and circle postulates.
These two initially contrasting themes continues to run strong in the eventual generalization to affine differential geometry.

\mbox{ } 

\ni Two furtherly primary types of geometry are, firstly, {\it ordering geometry}, which involves just a `intermediary point' variant of Euclid's line postulates. 
By involving {\sl neither} the parallel postulate nor the circle and right-angle pair of postulates, 
this can be seen as serving as a {\sl common} foundation for both absolute and affine geometry \cite{Coxeter}.
On the other hand, {\it projective geometry} involves ceasing to be able to distinguish between lines and circles in addition to angles being meaningless and no parallel postulate.
From a group-theoretic perspective, this is evoking the {\it projective linear group} $PGL(d, \mathbb{R}) = GL(d, \mathbb{R})/Z(GL(d, \mathbb{R}))$.\footnote{For a matrix group 
$\nFrG = GL(\nFrV)$ (for $\nFrV$ a vector space) or subgroups thereof, 
the {\it centre} $Z(\nFrG)$ of $\nFrG$ consists of whichever $k \, \mathbb{I}$ are allowed by the definition of that subgroup and the field $\mathbb{F}$ that $\nFrV$ is based upon.}

\mbox{ }

\ni Probably the best-known example of projective group is the {\it M\"{o}bius group} $PGL(2, \mathbb{C})$ acting upon $\mathbb{C} \cup \infty$ as the 
{\it fractional linear transformations} $z \longrightarrow \frac{az + b}{cz + d}$, for $a, b, c, d \, \in \, \mathbb{C}$ such that $ad - bc \neq 0$. 
This is 6-$d$, because there is one complex restriction on it: for $\lambda \in \mathbb{C}$, 

\ni\beq
\frac{\lambda \, a z + \lambda \,b}{\lambda \, c z + \lambda \, d} = \frac{\lambda}{\lambda} \, \frac{a z + b}{c z + d} = \frac{a z + b}{c z + d} \mbox{ } .  
\eeq
For practical use within Euclidean theories of space, note in particular that `spatial' measurements in our experience lie within the forms (\ref{Dist-Ratio}) and (\ref{Angle}), 
i.e. measuring tangible objects {\sl against a ruler} and measuring angles {\sl between} tangible entities.
On the other hand, more advanced, if indirect, physical applications make use of (extensions of) the other notions of geometry above.

\section{Lie groups and Lie algebras}\label{Lie}
%
A {\it Lie group} \cite{AMP} is simultaneously a group and a differentiable manifold; its composition and inverse operations are differentiable.  
Working with the corresponding infinitesimal (`tangent space') around $\FrG$'s identity element  -- the Lie algebra $\Frg$ -- 
is more straightforward due to vector spaces' tractability, whilst very little information is lost in doing so.
[E.g. the representations of $\Frg$ determine those of $\FrG$.]
More formally, a {\it Lie algebra} is a vector space equipped with a product (bilinear map) $\mbox{\bf |[} \mbox{ } \mbox{\bf ,} \mbox{ } \mbox{\bf ]|}: 
\Frg \times \Frg \longrightarrow \mbox{\Frg}$ that is antisymmetric and obeys the Leibniz (product) rule and the {\it Jacobi identity}
\beq
\mbox{\bf |[} g_1 \mbox{\bf ,} \, \mbox{\bf |[} g_2 \mbox{\bf ,} \, g_3 \mbox{\bf ]|} \, \mbox{\bf ]|} + \mbox{cycles} = 0 
\label{Jacobi-id}
\ee
$\forall \, g_1, g_2, g_3 \, \in \, \Frg$.
This is an example of {\it algebraic structure}: equipping a set with one or more product operations.  
In the present case, Lie brackets are exemplified by Poisson brackets and commutators.
Particular subcases of Lie brackets then include the familiar Poisson brackets and quantum commutators.

Moreover, a Lie algebra's generators (Lie group generating infinitesimal elements)  $\tau_{\sfp}$ obey 

\ni \beq
\mbox{\bf |[} \tau_{\sfp} \mbox{\bf ,} \, \tau_{\sfq} \mbox{\bf ]|} = {C^{\sfr}}_{\sfp\sfq}\tau_{\sfr} \mbox{ } ,
\label{Str-Const}
\eeq 
where ${C^{\sfr}}_{\sfp\sfq}$ are the {\it structure constants} of that Lie algebra.
It readily follows that the structure constants with all indices lowered are totally antisymmetric, and also obey  
\be
{C^{\sfo}}_{[\sfp\sfq}{C^{\sfr}}_{\sfs]\sfo} = 0  \mbox{ }  .  
\label{firstJac}
\ee  
%
Next suppose that it is hypothesized that some subset of the generators, $K_{\sfk}$, is significant.    
Denote the rest of the generators by $H_{\sfh}$.  
On now needs to check to what extent the algebraic structure in question actually complies with this assignation of significance.    
Such checks place limitations on how generalizable some intuitions and concepts which hold for simple examples of algebraic structures are.  
In general, the split algebraic structure is of the form

\ni\beq
\mbox{\bf |[} K_{\sfk} \mbox{\bf ,} \,  K_{\sfk^{\prime}} \mbox{\bf ]|} = {C^{\sfk^{\prime\prime}}}_{\sfk\sfk^{\prime}}K_{\sfk^{\prime\prime}} 
                                                                    + {C^{\sfh}}_{\sfk\sfk^{\prime}}               H_{\sfh}                 \mbox{ } ,
																	\label{Lie-Split-1}
\eeq 

\ni\beq
\mbox{\bf |[} K_{\sfk} \mbox{\bf ,} \,  H_{\sfh}          \mbox{\bf ]|} = {C^{\sfk^{\prime}}}_{\sfk\sfh}K_{\sfk^{\prime}} 
                                                                    + {C^{\sfh^{\prime}}}_{\sfk\sfh}H_{\sfh^{\prime}}                       \mbox{ } ,
																	\label{Lie-Split-2}
\eeq 

\ni\beq
\mbox{\bf |[} H_{\sfh} \mbox{\bf ,} \,  H_{\sfh^{\prime}} \mbox{\bf ]|} =  {C^{\sfk}}_{\sfh\sfh^{\prime}}               K_{\sfk}                
                                                                    +  {C^{\sfh^{\prime\prime}}}_{\sfh\sfh^{\prime}}H_{\sfh^{\prime\prime}} \mbox{ } .
																	\label{Lie-Split-3}
\eeq 
Denote the second to fifth right hand side terms by 1) to 4).  
1) and 4) being zero are clearly subgroup closure conditions.
2) and 3) are `interactions between' $\Frh$ and $\Frk$.
The following cases of this are then realized in this paper.  

\mbox{ } 

\ni I)   {\it Direct product}. If 1) to 4) are zero,   then $\Frg = \Frk  \times \Frh$. 

\ni II)  {\it Semi-direct product}. If 2) alone is nonzero, then $\Frg = \Frk \rtimes \Frh$. 

\ni III) {\it Thomas integrability}. If 1) is nonzero, then $\Frk$ is not a subalgebra: attempting to close it leads to some $K_{\sfk}$ are discovered to be integrabilities.
I denote this by $\Frk \mbox{\textcircled{$\rightarrow$}} \Frh$; the arrow points to the part of the split which arises as an integrability of the other part.  
A simple example of this occurs in splitting the Lorentz group's generators up into rotations and boosts; 
this is indeed the group-theoretic underpinning \cite{Gilmore} of Thomas precession (see Appendix B.1).

\ni IV) {\it Two-way integrability} If 1) and 4) are nonzero, neither $\Frk$ nor $\Frh$ are subalgebras, due to their imposing integrabilities on each other.
I denote this by $\Frk \mbox{\textcircled{$\leftrightarrow$}} \Frh$, with the double arrow indicating that the two parts of the split are integrabilities of each other.  
In this case, any wishes for $\Frk$ to play a significant role by itself are almost certainly dashed by the mathematical reality of the algebraic structure in question.

\mbox{ } 

\ni Note that this classification is important as regards understanding how GR's constraints are more subtle than Gauge Theory's, and Supergravity's than GR's; 
this is further developed in Appendices C and D.

\subsection{Examples of Lie groups and Lie algebras}\label{Lie-Ex} 

For Abelian Lie groups, the structure constants are all zero. 
Examples of this include $Tr(d)$ and $Dil(d)$ -- which are 

\ni both noncompact -- and the compact $Rot(2) = SO(2)$. 
The corresponding Lie algebras' generators are $P_{i}  = - \frac{\pa}{\pa x^i}$, $D =  - x^{i}\frac{\pa}{\pa x^i}$ and $L = y \frac{\pa}{\pa x} - x \frac{\pa}{\pa y}$.

On the other hand, for $d > 2$ the $Rot(d) = SO(d)$ are non-Abelian Lie groups.
Compare these with the $O(d)$ Lie groups: $SO(d)$ is a Lie subgroup of $O(d)$. 
The former has 2 connected components related by a discrete reflection. 
The corresponding Lie algebra sees only the connected component that contains the identity, so is the same for each of $O(d)$ and $SO(d)$.
$GL(\FrV)$ and $SL(\FrV)$ are also Lie groups [non-Abelian for $\mbox{dim}(\FrV) > 1$].

Some Lie algebras used in this paper are as follows. 
The {\it general linear algebras} $gl(\FrV)$ of $d \times d$ matrices over $\mathbb{F}$, the real cases of which have dimension $d^2$. 
The {\it special linear algebras}     $sl(\FrV)$ are the zero-trace such,                the real cases of which have dimension $d^2 - 1$.
The generators for $gl(d, \mathbb{R})$ are, very straightforwardly, $G_{ij} = x^{i}\frac{\pa}{\pa x^{j}}$.  
The {\it special orthogonal algebras} $so(d) := \{A \in gl(d, \mathbb{R}) \, | \, A + A^{\sT} = 0 \}$            of dimension $d\{d - 1\}/2$. 
These are generated by $M_{ij} = x^{i}\frac{\pa}{\pa x^{j}} - x^{j}\frac{\pa}{\pa x^{i}}$ subject to the schematic noncommutation relation
\beq 
\mbox{\bf |[}M\mbox{\bf ,} \, M\mbox{\bf ]|} \sim M \mbox{ } .
\label{M-M}
\eeq
Among these, $so(2)$ is the above-mentioned Abelian algebra and $so(3)$ has the alternating symbol ${\epsilon^i}_{jk}$ for its structure constants.
The 3-$d$ case also simplifies by the duality between $M_{ij}$ and $L_i$. 
$SO(4)$ [and, more famously, $SO(3, 1)$: the Lorentz group] satisfy accidental relations linking them to a direct product of two copies of $SO(3)$.
Moreover, in the $SO(3, 1)$ case 
-- whereas linear combinations can be taken so as to obtain this split, the original presentation's generators differ in physical significance (3 rotations and 3 boosts) -- 
which physical meanings are not preserved by taking said linear combinations. 
Adhering then to the physically meaningful split into $J_i$ and $K_i$ is the setting of the Thomas precession mentioned in the previous sub-appendix. 
Schematically, this decomposes (\ref{M-M}) into 
\beq
\mbox{\bf |[}J \mbox{\bf ,}  \, J\mbox{\bf ]|} \sim J     \mbox{ } , \mbox{ } \mbox{ } 
\mbox{\bf |[}J \mbox{\bf ,}  \, K\mbox{\bf ]|} \sim K     \mbox{ } , \mbox{ } \mbox{ } 
\mbox{\bf |[}K \mbox{\bf ,}  \, K\mbox{\bf ]|} \sim K + J \mbox{ } , 
\eeq
the key bracket being the last one by which the boosts are not a subalgebra, the precession in question referring to the rotation arising thus from a combination of boosts.

\mbox{ }

\ni Some composite Lie groups of particular relevance to this paper are then $Eucl(d)$ and $Sim(d)$.
Here e.g. $Eucl(d)$'s semidirect product structure is due to the bracket 
$\mbox{\bf |[} J \mbox{\bf ,} \, P \mbox{\bf |]} \sim P$, signifying that $P$ is a $Rot(d)$-vector.
Also, $Tr(d) \rtimes Dil$'s semidirect product structure rests upon $\mbox{\bf |[} P \mbox{\bf ,} \, D \mbox{\bf ]|} \sim P$.  
On the other hand, $Rot(d)$--$Dil$ independence as found e.g. within the family of subgroups of $Aff(d)$ [and thus in particular $Eucl(d)$ and $Sim(d)$] 
is based upon rotation and dilation generators commuting: a direct product split

\ni\beq
\mbox{\bf |[} M \mbox{\bf ,} \, D \mbox{\bf ]|}             = 0 \mbox{ } \mbox{ } (            \mbox{ } \mbox{ }
\mbox{\bf |[} \underline{L} \mbox{\bf ,} \, D \mbox{\bf ]|} = 0 \mbox{ } \mbox{ in 3-$d$ } \mbox{ and } \mbox{ } \mbox{ } 
\mbox{\bf |[} L \mbox{\bf ,} \, D \mbox{\bf ]|}             = 0 \mbox{ } \mbox{ in 2-$d$ }) .
\eeq

\subsection{Killing vectors and isometries: plain, homothetic and conformal}\label{KV}
%
{            \begin{figure}[ht]
\centering
\includegraphics[width=0.25\textwidth]{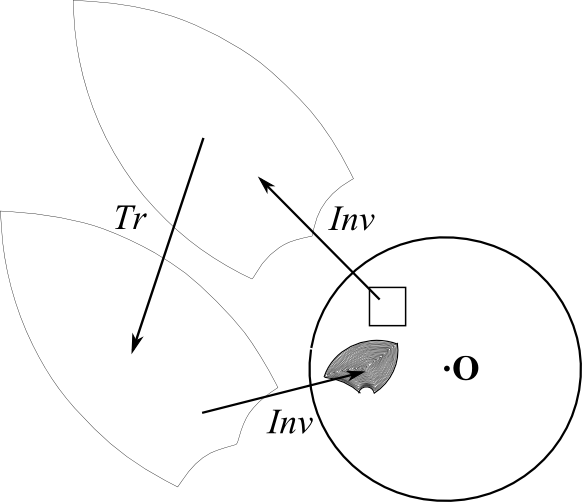}
\caption[Text der im Bilderverzeichnis auftaucht]{        \footnotesize{Decomposition of special conformal transformation into an inversion, translation and another inversion.} }
\label{More-G-Lie} \end{figure}          }

\ni One way of getting at $Eucl(d)$ which usefully extends to further groups is by solving Killing's equation in flat space.
In this case, given Killing's Lemma \cite{Wald}, the form 
\be
\upxi^{i} = a^{i} + {B^{i}}_{j}x^{j}
\ee
for the Killing vectors readily follows.
Repeating for the homothetic Killing equation in flat space, 
\beq 
\upxi^{i} = a^{i} + {B^{i}}_{j}x^{j} + c x^{i}  
\label{CK-Sim}
\eeq
ensues.
Finally, repeating for the conformal Killing equation in flat space  
\beq 
\upxi^{i} = a^{i} + {B^{i}}_{j}x^{j} + c x^{i} + \{ 2 k^{j} x^{i} - k^{i} x^{j} \} \delta_{ik} x^{k} \mbox{ } . 
\label{CK-Conf}
\eeq
ensues for $d \geq 3$ \cite{AMP}. 
The $k^i$ correspond to {\it special conformal transformations} 

\ni\beq
x^i \longrightarrow \frac{x^i - k^i x^2}{1 - 2 \underline{k}\cdot\underline{x} + k^2 x^2}
\eeq
formed from an inversion, a translation and then a second inversion. 
The infinitesimal generator is 

\ni $K_i := x^2\frac{\pa}{\pa x^i} - 2 x_i x^j \frac{\pa}{\pa x^j}$.  
Thus conformal group $Conf(d)$ of dimension $d\{d + 2\}/2$ (in particular 10 for $d = 3$) arises.

For $d = 2$, the conformal Killing equation famously collapses to the  Cauchy--Riemann equations, causing an infinity of solutions: any holomorphic function $f(z)$ will do.
In 1-$d$, the conformal Killing equation collapses to $\d \xi/\d x = \phi(x)$, 
amounting to reparametrization by a 1-$d$ coordinate transformation $v = \Phi(x) + a$ for $\Phi := \int\phi(x)\d x$. 
This case is {\sl not} subsumed within (\ref{CK-Conf}): $Conf(1)$ is also infinite-dimensional, albeit rather less interesting than the 2-$d$ version.   
(1-$d$ has no angles to preserve, though conformal factors can be defined for it none the less; note also that the metric {\sl drops out} of the 1-$d$ conformal Killing equation.)

Due to an integrability of the schematic form 

\ni\beq
\mbox{\bf |[}  K  \mbox{\bf ,} \,  P  \mbox{\bf ]|} \sim M + D \mbox{ } ,
\label{K-P}
\eeq 
the conformal algebra is $(P, K) \mbox{\textcircled{$\rightarrow$}} (M, D)$ Thomas.
I.e. a translation and an inverted translation compose to give both a rotation (`conformal precession') {\sl and} an overall expansion.
Elsewise $K_i$ behaves much like $P_i$ does.

\subsection{Some further subgroups acting upon $\mathbb{R}^d$}\label{+G}

The above three Sections can be viewed as introducing $P_i$, $M_{ij}$, $D$ and $K_{i}$ generators.

One can furthermore consider e.g. shears and $d$-dimensional volume preserving stretches ($G_{(ij)}^{\sT}$ generators); each of these are only nontrivial for $d \geq 2$.   
Alongside the rotations, these form $SL(d, \mathbb{R})$; 
then the {\it equi-top-form group} $E(d) := Tr(d) \rtimes SL(d, \mathbb{R})$, corresponding to the eponymous geometry (equiareal in 2-$d$ \cite{Coxeter}).
Also, $GL(d, \mathbb{R}) = Dil \times SL(d, \mathbb{R})$; then the {\it affine group} $E(d) := Tr(d) \rtimes GL(d, \mathbb{R})$, corresponding to affine geometry \cite{Coxeter}.  
dim($E(d)$) = $d\{d + 1\} - 1$ and dim($A(d)$) = $d\{d + 1\}$. 
The unsplit nonzero affine brackets are, schematically, 
\beq
\mbox{\bf [} G \mbox{\bf ,} \, G \mbox{\bf ]} \sim G \mbox{ } , \mbox{ } \mbox{ } \mbox{\bf [} G \mbox{\bf ,} \, P \mbox{\bf ]} \sim P \mbox{ } , 
\eeq
signifying closure of the $GL(d, \mathbb{R})$ subgroup and that $P_i$ is a  $GL(d, \mathbb{R})$  vector.  
As regards general $sl(d, \mathbb{R})$ generators, perform the antisymmetric--symmetric and trace--tracefree splits on $gl(d, \mathbb{R})$'s generators (Fig \ref{Flat-Geom}). 
Then the antisymmetric part is just the rotations, the tracefree symmetric part is 
$E_{ij} = x^{i}\frac{\pa}{\pa x^{j}} + x^{j}\frac{\pa}{\pa x^{i}} - \frac{2}{n} \, {\delta^i}_j \, x^{k}\frac{\pa}{\pa x^{k}}$ and the trace part (the usual dilation) is discarded.
E.g. Corresponding infinitesimal matrices for $sl(2, \mathbb{R})$ these are $\left(\stackrel{1 \mbox{ } \, 0}{\mbox{\scriptsize 0         --1}}\right)$ 
                                                            and $\left(\stackrel{0 \mbox{ } \, \, 1}{\mbox{\scriptsize 1 \mbox{ } 0}}\right)$ which form the triple (\ref{SL-Array}) 
with the infinitesimal rotation matrix.                   
The corresponding generators are $= x\frac{\pa}{\pa x} - y\frac{\pa}{\pa y}$ for Procrustean stretches and $= x\frac{\pa}{\pa y} + y\frac{\pa}{\pa x}$ for shears.  
It is also important to note that 

\ni\beq
\mbox{\bf [}\mbox{Shear}\mbox{\bf ,} \, \mbox{Shear}^{\prime}\mbox{\bf ]} \sim \mbox{Rotation} 
\label{Sh-Sh} 
\eeq
by which the non-rotational parts of $SL(d, \mathbb{R})$ cannot be included in the absense of the rotations.
%
{            \begin{figure}[ht]
\centering
\includegraphics[width=0.9\textwidth]{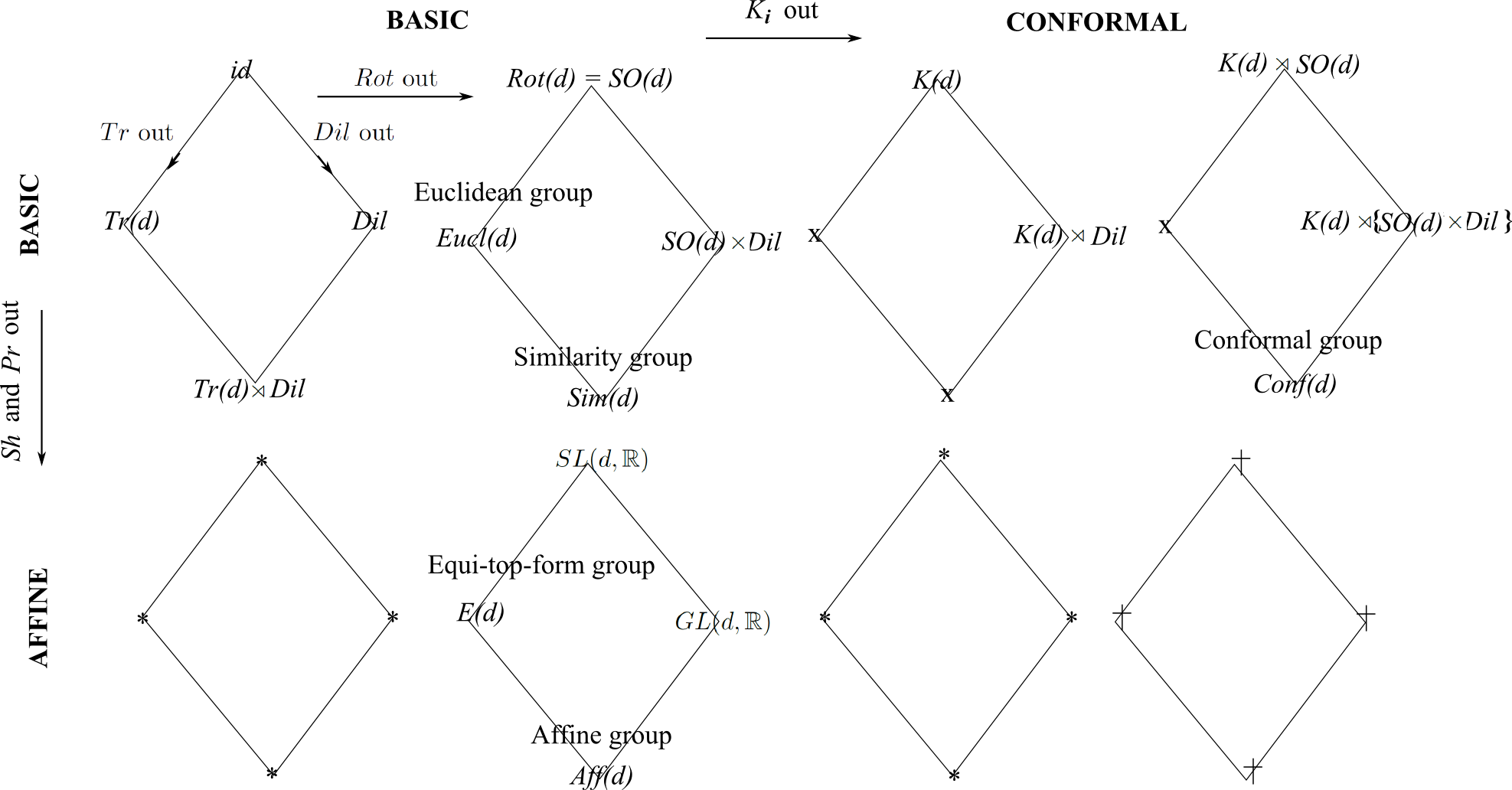}
\caption[Text der im Bilderverzeichnis auftaucht]{        \footnotesize{Summary sketch, of groups including further groups acting upon $\mathbb{R}^d$. 
These are arrived at by adding generators as per the labelled arrows.
Moreover, the group relations involved do not permit all combinations of generators to be included.
In particular, absences marked $X$ are due to integrability     (\ref{K-P}).
Absences marked $*$ are due to integrability     (\ref{Sh-Sh}).
Finally, absences marked $\dagger$ are due to obstruction (\ref{K-S}).
Figures \ref{More-G-Invariants} and \ref{More-G} then use a matching layout. } }
\label{More-G-Lie-2} \end{figure}          }
 
Moreover, the $K_i$ and $G_{(ij)}^{\sT}$ generators are not compatible with each other, as is clear from 
\beq
\mbox{conformal transformations only preserving angles whereas shears do not preserve angles } .
\label{K-S}  
\eeq
Thus there are two distinct `apex groups': $Conf(d)$ from including $K_i$ and $Aff(d)$ from including $S_{ij}$.
`Apex' is used here in the sense that the other possibilities are contained within as Lie subgroups.
These include a number of subgroups not yet considered (Fig \ref{More-G-Lie}).

Finally, some of the above groups also support nontrivially distinct projective versions, obtained by quotienting out by the centre of the group in question. 
E.g. if this is performed upon the affine group, projective geometry \cite{Coxeter, Stillwell} ensues.

\subsection{Superconformal and superaffine algebras and some of their subalgebras}\label{Susy}

Some algebraic structures involve {\it anticommutators} $\mbox{\bf |[} A \mbox{\bf ,} \, B \mbox{\bf ]|$_{+}$} := A B + B A$
These enter models of fermionic species.   
See e.g. \cite{PSBook, Frenkel} for an extensive `mathematical methods for physicists' treatments of these and of the ensuing notion of spinors.

As compared to the conformal group $Conf(d)$, the superconformal group $superConf(d)$ \cite{Weinberg3} has additional fermionic generators $S$ and $Q$ (and conjugates), 
in which sense it is doubly supersymmetric (denoted $\mN = 2$). 
Here, 
\beq
\mbox{\bf |[}S\mbox{\bf ,} \, \bar{S}\mbox{\bf ]|$_+$} \sim P \mbox{ } 
\label{S-S}
\eeq
and 
\beq
\mbox{\bf |[}Q\mbox{\bf ,} \, Q\mbox{\bf ]|$_+$} \sim K \mbox{ } ,
\label{Q-Q}
\eeq
so both the momentum and the special conformal transformation arise as integrabilities.

On the other hand, the superaffine algebra $superAff(d)$ has just the one additional fermionic generator $S$ (and conjugate).
Then (\ref{S-S}) holds, by which $superAff(d)$ is $S \mbox{\textcircled{$\rightarrow$}} (P, G)$ Thomas (composition of supersymmetries results in a translation).
Examining the table of subgroups in Fig \ref{Susy-Groups}, the exhibited supersymmetric subgroups of $superAff(d)$ share this property.
In contrast, the superconformal group has a more complicated sequence of integrabilities, 
with $(S, Q) \mbox{\textcircled{$\rightarrow$}} (P, K) \mbox{\textcircled{$\rightarrow$}} (M, D)$; this is a tripartition to (\ref{Lie-Split-1}-\ref{Lie-Split-3})'s bipartition.
Finally note that all the groups in the table bar $superConf(d)$ have just the one supersymmetric generator (and conjugate): $\mN = 1$.  
8 are supersymmetric subgroups of $superConf(d)$ and 5 of $superAff(d)$ [4 of which are also among the previous 8].
%
{            \begin{figure}[ht]
\centering
\includegraphics[width=0.9\textwidth]{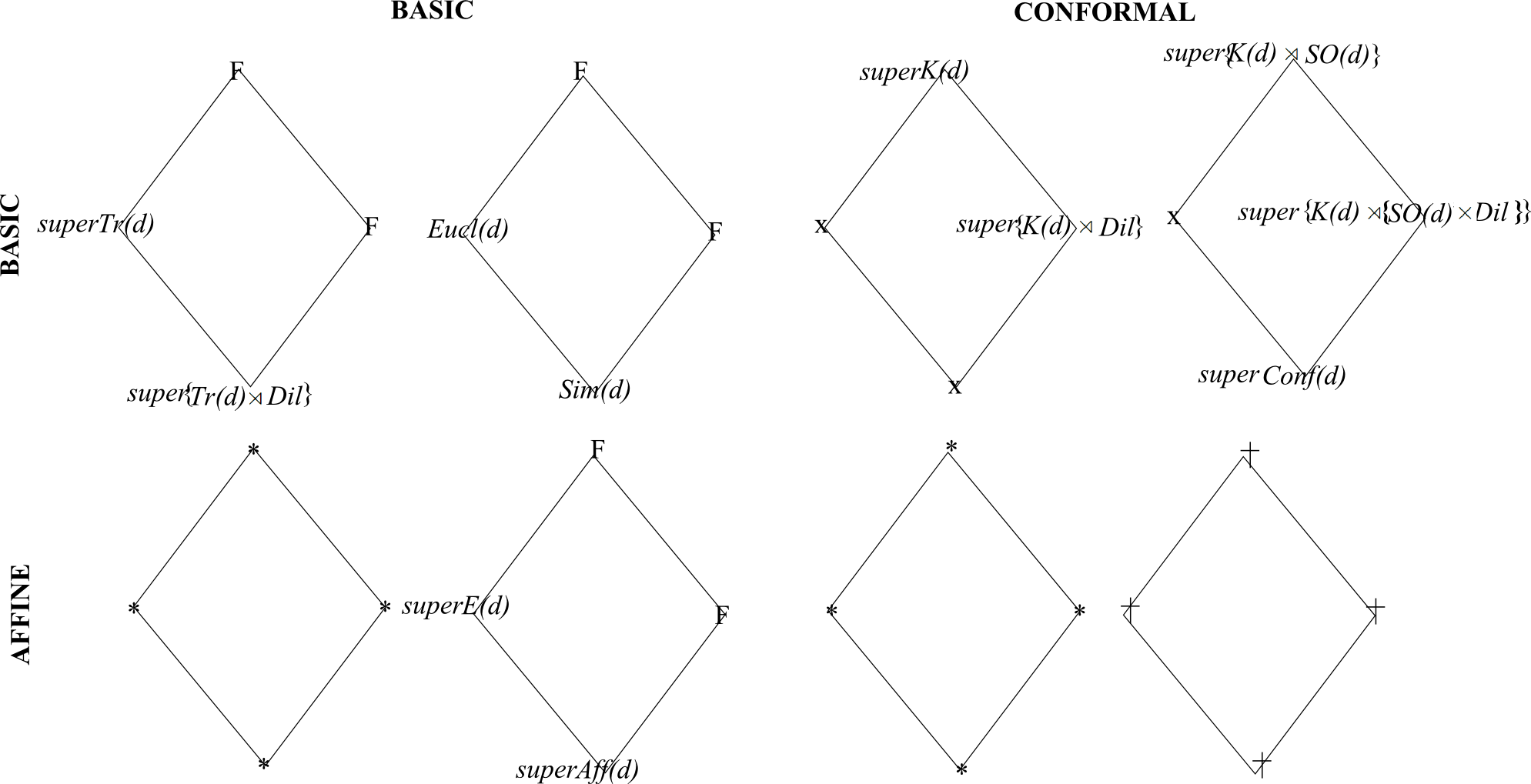}
\caption[Text der im Bilderverzeichnis auftaucht]{        \footnotesize{Some subgroups, and overruled combinations, of supersymmetric groups, 
within the superconformal and superaffine apex groups.
These are again laid out according to \ref{More-G}, except now with at least one supersymmetric generator as well.  
The cases marked with an F fail to support Supersymmetry due to Supersymmetry requiring at least one of $P_i$ and $K_i$ as an integrability: (\ref{S-S}-\ref{Q-Q}).} }
\label{Susy-Groups} \end{figure}          }

\section{Flat to differential geometry modelling of space, and GR-level counterpart of this paper}\label{Diff-Geom}
%
{            \begin{figure}[ht]
\centering
\includegraphics[width=0.8\textwidth]{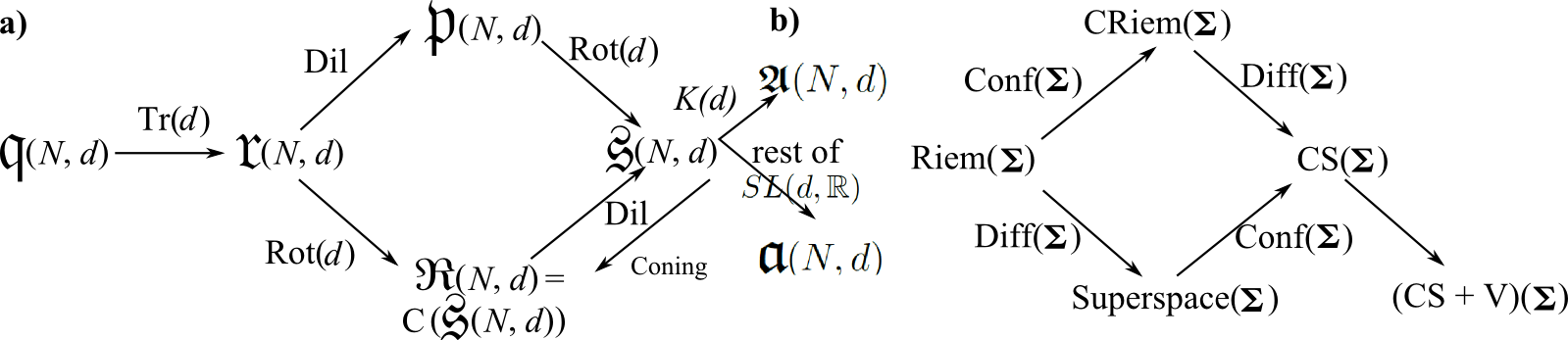}
\caption[Text der im Bilderverzeichnis auftaucht]{        \footnotesize{a) This Sec's specific sequence of configuration spaces, as a useful model arena for GR's outlined in b).  
CS denotes conformal superspace and $\mV$ denotes a solitary global volume degree of freedom.} }
\label{Q-RPM-GR} \end{figure}          }

\ni A) Firstly, metric geometry is one of the outcomes of weakening the axioms of Euclidean geometry.  
In particular, a first route to such is through assuming only absolute geometry and then finding a large multiplicity of such to exist. 
Moreover, some aspects of `metric geometry' remain upon ceasing to assume a metric. 
For instance, some such aspects form topological manifold geometry and differentiable manifold geometry. 
Furthermore, many of the outcomes of stripping down the structure of Euclidean geometry have close counterparts in differential geometry. 
E.g., similarity, conformal, affine and projective {\sl differential} geometries exist \cite{KN, Kobayashi}.
Additionally, it is possible to reconcile e.g. affine and metric (or conformal metric) notions in differential geometry. 
E.g. the standard modelling assumptions in GR include that the metric connection serve as a (and the only) affine connection.

\mbox{ } 

\ni In this way, stripping away the layers of structure assumed in GR (whether to model spacetime, or to model space within a geometrodynamical perspective) 
leads to closely analogous first ports of call to those in stripping away the layers of structure assumed in RPM's based upon $\mathbb{R}^d$.  
In the specific case of GR, these first ports of call \cite{AGates} are conformal geometry and affine geometry. 
Upon Supersymmetry's extra structure, Supergravity is another first port of call.    
In the case of studying GR configurations, physical irrelevance of Diff($\bupSigma$) is also under consideration; 
the most common corresponding configuration spaces are depicted in Fig \ref{Q-RPM-GR}.b).  
Here $\bupSigma$ some spatial topology taken in the current paper to have some fixed compact without boundary form.  
The Figure juxtaposes these with analogous RPM configuration spaces as laid out in the main part of the current paper.
One can also consider an analogous diamond of affine spaces, 
with variants on GR allowing for the additional possibility of both retaining a metric and introducing an affine connection other than the metric connection. 
Indeed, one means of such an alternative theory having torsion is through the difference of two distinct affine connections constituting a torsion tensor.
Superconformal Supergravity is an example of a further combination of the elements of these first ports of call. 
All of these are open to relational analyses and yet wider consideration of Background Independence leading to whether they exhibit significant differences from GR 
as regards the Problem of Time's many facets.

A {\sl second} port of call is considering Topological Relationalism 
(whether in the context of variants on GR allowing for topology change, or in the context of Topological Field Theories being a further generalization of Conformal Field Theories).
See \cite{ASoS} for discussion of the second {\sl and subsequent} ports of call from a relational perspective.    
What about modelling this second port of call with RPMs?  
In fact, this is relatively straightforward. 
E.g. in one sense {\it change in particle number} in RPMs is analogous to topology change in GR. 
(I.e. theories which {\sl retain} the upper layers of structure whilst letting subsequent layers of structure also be dynamical rather than absolute backgrounds). 
In another sense (stripping away the upper layers), topological RPMs involve distinguishing only those configurations which are topologically distinct. 
Then e.g. scaled triangleland collapses to a small finite number of points: 
the total collision, the double collision (or three such if labelled) and the general configuration that is not any of the previous.
This is rather simpler than the space of all topological manifolds in some given dimension! 
Thus it is very plausible for RPMs to be able to model multiple layers of structure (at the cost of resembling GR rather less at the lower levels).  

\mbox{ }

\ni B) Let us next return to the upper layers of structure, so as to justify some of the many GR--RPM analogies that occur there. 
GR can be cast in Temporal and Configurational Relationalism form \cite{RWR}, the latter being tied to the group of spatial diffeomorphisms $Diff(\bupSigma)$.
The underlying configuration space is $\mbox{Riem}(\bupSigma)$, i.e. the space of spatial (positive-definite) 3-metrics on $\bupSigma$.  
Here the constraint provided by Temporal Relationalism \cite{RWR} gives 
-- via Dirac's argument that reparametrization invariance necessarily implies a primary constraint \cite{Dirac} -- 
a relational recovery of the quadratic {\it GR Hamiltonian constraint} ${\cal H}$ \cite{ADM}.   
On the other hand, Configurational Relationalism \cite{RWR, FileR} provides the linear {\it GR momentum constraint} ${\cal M}_i$ \cite{ADM}. 
In this sense, Temporal and Configurational Relationalism remain distinct Background Independence aspects in GR, much as they also are in RPMs.  
Each provides a constraint of its own.

These constraints then close in accord with the {\it Dirac algebroid} of GR, which is schematically of the form
\beq
\mbox{\bf \{} {\cal M} \mbox{\bf ,} \, {\cal M} \mbox{\bf \}} \sim {\cal M} \mbox{ } , \mbox{ } \mbox{ } 
\eeq
\beq
\mbox{\bf \{} {\cal M} \mbox{\bf ,} \, {\cal H} \mbox{\bf \}} \sim {\cal H} \mbox{ } , \mbox{ } \mbox{ } 
\eeq
\beq
\mbox{\bf \{} {\cal H} \mbox{\bf ,} \, {\cal H} \mbox{\bf \}} \sim {\cal M} \mbox{ } . 
\label{Dirac-Algebroid}
\eeq
These are the classical theory's Poisson brackets; this closure is a classical realization of the further Constraint Closure aspect of Background Independence.

The first bracket means that the ${\cal M}_i$ -- which correspond to the spatial diffeomorphisms $Diff(\bupSigma)$ -- themselves close to form a true infinite-$d$ Lie algebra. 
The second bracket signifies that ${\cal H}$ is a $Diff(\bupSigma)$ scalar density. 
The third bracket is not only the one expressing an integrability \cite{MT72} but also the one containing both a structure function and the derivative of its RHS constraint.
Its integrability means that in GR, Temporal Relationalism obliges the existence of nontrivial Configurational Relationalism. 
This feature does not occur in RPMs, in which Temporal and Configurational Relationalism can each be modelled in the absense of the other.  
The third bracket can furthermore be viewed as an algebroid\footnote{An {\it algebroid} allows for `structure functions' -- including derivative operators in the present case --
of constraints to appear on the right-hand side.}  
counterpart of Thomas precession: ${\cal H} \mbox{\textcircled{$\rightarrow$}} {\cal M}_i$.  
In more detail, compare how 
composing two      boosts                                                                                          results in a rotation:                        Thomas precession,  
whereas composing  two time evolutions -- or pure hypersurface deformations in \cite{HKT}'s interpretation of ${\cal H}$ -- results in a spatial diffeomorphism: Moncrief--Teitelboim 
                                                                                                                                                                 on-slice Lie dragging. 
[Lie dragging is the motion corresponding to Diff($\bupSigma$) in the same manner as precession is a name for a motion corresponding to $Rot(d)$.]
This analogy is (as far as the Author is aware) new to this paper.  
As well as rendering the Dirac Algebroid more pedagogically accessible to students, 
this analogy is useful in the below revelations about Supergravity being substantially different from GR 
as regards the form taken by its Relationalism and Background Independence more generally.

Further consequences of the above algebraic form as regards background independent features are as follows. 

\mbox{ }

\ni 1) that Wheeler's $\mbox{Superspace}($\bupSigma$) := \mbox{Riem}($\bupSigma$)/Diff($\bupSigma$)$ -- central to geometrodynamics --
is a meaningful intermediary configuration space.
This follows from ${\cal M}_i$ forming a subalgebra of the Dirac algebroid by the first bracket, so it is meaningful to reduce out ${\cal M}_i$ by itself.  

\ni 2) Expression in terms of Beables or Observables is a fourth aspect of Background Independence; 
the general failure of which in classical and quantum gravitational theories then constitutes the well-known `Problem of Observables' facet of the Problem of Time.
Observables or beables are objects which commute with constraints.
In the case of commuting with all constraints -- ${\cal H}$ and ${\cal M}_i$ in GR -- one is dealing with Dirac observables. 
In the case of commuting with linear constraints only -- ${\cal M}_i$ in GR -- one is dealing with \K observables \cite{Kuchar93}.  
Moreover, concepts of observables or beables as objects which commute with subsets of constraints only make sense if the subset in question algebraically closes \cite{ABeables}. 
Thus ${\cal M}_i$ forming a subalgebra of the Dirac algebroid by the first bracket guarantees the meaningfulness of \K observables in the case of GR.  
See \cite{ABeables-2} for the form taken by the \K observables for the current paper's new non-supersymmetric theories.  

\mbox{ }

\ni Further conformal Configurational Relationalism has also been considered in the case of GR and of alternative theories of conformogeometrodynamics \cite{ABFO, GR-Config2, ABFKO-SD}.
The original conception of this used metric shape RPM as a model arena; however, the current paper's conformal shape RPM is surely a comparable or enhanced source of insights.

\mbox{ }

\ni \underline{Frontier 15}. Finally, returning to topic A), 
perform the differentiable manifold level counterpart of the current  paper's comparative relational study of the diverse levels of structure to be found in flat geometries.

\mbox{ } 

\ni The current paper and Frontier 15 can furthermore be seen as part of a wider program in which an increasing number of levels of mathematical structure assumed in Physics 
are taken to be dynamical rather than fixed background structures.
This program was initiated by Isham \cite{Ishamic}, 
who considered replacing geometrical quantization based upon the usual configuration space with that based upon generalized configuration spaces.
I then provided a classical precursor for this program in \cite{ASoS}, in particular sketching out an even wider range of classical Shape Statistics theories than the current paper's 
(at the levels of topological manifolds and of topological spaces).  
That work also posed Shape Statistics on GR's Superspace($\bupSigma$) and conformal superspace, CS($\bupSigma$).
Those remaining a distant dream in the general case, 
I also posed the more tractable analogues in the settings of anisotropic minisuperspace and of inhomogeneous perturbations about minisuperspace \cite{AConfig, ABook}.

\section{Contrast with canonical formulation of Supergravity}\label{Sugra}

Supersymmetry and split space-time GR can each separately be envisaged as Thomas-type effects at the level of algebraic structures.
Furthermore, upon considering both at once, the integrabilities involved go in opposite directions, forming the more complicated `two-way' integrability case). 
[Contrast also with the superconformal group's composition of integrabilities being tripartite with two aligned one-way integrabilities instead.]
The schematic form of the key new relation for Supergravity is (see e.g. \cite{VM10})  
\beq
\mbox{\bf \{}{\cal S}\mbox{\bf ,}\, {\cal S}\mbox{\bf \}}_{\sC} \sim {\cal H} + {\cal M} \mbox{ } . \mbox{ } \mbox{ }  
\label{S-S->H}
\eeq
This forms the second integrability of the `two-way' pair, to the first integrability being of form (\ref{Dirac-Algebroid}).\footnote{The C subscript stands for Casalbuoni Poisson bracket 
\cite{Casalbuoni}. 
The time component $P_0$ arising within (\ref{S-S}) in the indefinite case's superPoincar\'{e} subgroup can be seen as a precursor of (\ref{S-S->H}).}

\mbox{ } 

\ni The implications of the two-way integrability case include that the linear constraints do not close by themselves; thus

\mbox{ } 

\ni 1) they cannot be quotiented out as a unit (in this sense {\sl no} Supergravity counterpart of Wheeler's Superspace).

\ni 2) Temporal and Configurational Relationalsim become a fused notion as opposed to separate notions.  

\ni 2) There is {\sl no} Supergravity counterpart of \K observables.

\mbox{ } 

\ni Moreover, two more possibilities for splits manifest themselves. 
The main idea here is that the Temporal to Configurational Relationalism split, 
the quadratic and linear constraints split and the notion of \K observables correspond to treating the linear consraints ${\cal LIN}$ differently in GR.
However, in a wider range of theories including Supergravity, this consideration is to be supplanted more generally by splits which respect the subalgebraic structures 
contained within the constraint algebraic structure.

\mbox{ }

\ni Then as a second possibility for a split \cite{AGates}, one can also consider the ${\cal S}, {\cal H}$ to ${\cal LIN}^{\prime}$ split (for ${\cal LIN}^{\prime}$ 
the linear non-supersymmetric constraints); this is Thomas with a ${\cal LIN}^{\prime}$ subalgebra.  

\mbox{ } 

\ni As a third possibility, the ${\cal LIN}^{\prime}$, ${\cal H}$ to ${\cal S}$ split is Thomas, exhibiting a `non-supersymmetric' subalgebraic structure.
In particular, these other splits permit a meaningful notion of quotienting out ${\cal LIN}^{\prime}$, 
giving a well-defined quotient space and a well defined notion of observables in this modified sense.
I.e. 

\mbox{ } 

\ni 1.A) extend Riem($\bupSigma$) to include the space of gravitino fields, and then quotient out solely by the non-supersymmetric linear constraints. 

\ni 2.A) A notion of observables or beables can be defined as commuting with solely the non-supersymmetric linear constraints.

\mbox{ } 

\ni The yet further versions 1.B) and 2.B) removing the word `linear' from the preceding definitions are likely to be harder to handle.
I term 2) and 2.A) {\it non-supersymmetric \K and Dirac observables or beables} respectively. 
I term 1) and 1.A) {\it non-supersymmetric Superspace and True-space} respectively; True-space is a formal reference to the space of true dynamical degrees of freedom itself.
Finally, whether any of the entities termed `non-supersymmetric' 
are in violation of the spirit of Supersymmetry may be a matter further relevance from viewpoints which take Supersymmetry sufficiently seriously.  

\mbox{ } 

\ni Furthermore, due to ${\cal H}$'s ties to Temporal Relationalism in GR, and of (some) linear constraints' ties to Configurational Relationalism, 
Supergravity's change of status as regards which of these constraints can be entertained independently of which others also concerns how Relationalism is to be viewed.
Additionally, Temporal Relationalism, Configurational Relationalism, Constraint Closure and Expression in terms of Beables are four of the aspects of Background Independence \cite{APoT3}
underlying four of the Problem of Time facets \cite{Kuchar92, I93, APoTAPoT2}. 
Thus Supergravity exhibits a very different realization of these from GR \cite{ABeables}, 
which could herald one or both of foundational problems for Supersymmetry or a hitherto insufficiently general conceptualization of Background Independence and the Problem of Time.  
One possibility here is that Supersymmetry breaks down the divide between Temporal and Configurational Relationalism as separate providers of constraints. 
Another possibility is that Supersymmetry renders constraint provision tripartite, by itself constituting a third provider of constraints.  
The current  paper then shows that {\sl none} of the above happen in superRPMs whose supersymmetry is tied to the translations. 
In this case, Supersymmetry is {\sl compatible} with Relationalism and is implemented as a subcase of Configurational Relationalism.  
Of course, the current  paper also points out that superRPMs lack an analogue of (\ref{Dirac-Algebroid}). 
In this manner they realize the {\sl opposite} single plank to that realized by non-supersymmetric GR. 
It is then rather interesting that {\sl the existing conception of Background Independence can be combined with Supersymmetry without requiring modification of either}, 
at least in these nontrivial and complementary model arenas.

\mbox{ }

\ni \underline{Frontier 16}. Resolve the above matter in full Supergravity. 
Or at least do so in some model arena exhibiting specifically the `two-way' pair of integrabilities, by which ${\cal S}$ implies ${\cal H}$, and ${\cal H}$ imply ${\cal LIN}^{\prime}$, 
with the Temporal to Configurational Relationalism divide hitherto having been between ${\cal H}$ and the full set of linear constraints.
Does this have any further consequences for the `space and configuration primality' versus `spacetime primality' debate? \cite{Battelle, RWR, APoT3}.  

\end{appendices}


\end{document}